\documentclass[prb,showpacs,superscriptaddress,twocolumn,amssymb,amsfonts]{revtex4}
\usepackage{amsmath}
\usepackage{bm}
\usepackage{epsfig}
\usepackage{graphics}
\usepackage{dsfont}

\def\beq{\begin{equation}}
\def\eeq{\end{equation}}
\def\beqn{\begin{eqnarray}}
\def\eeqn{\end{eqnarray}}

\def\E{{\bf E}}
\def\B{{\bf B}}
\def\P{{\bf P}}
\def\M{{\bf M}}

\def\k{{\bf k}}
\def\r{{\bf r}}
\def\q{{\bf q}}
\def\v{{\bf v}}
\def\cc{\mathrm{c.c.}}

\def\F{{\cal F}}
\def\A{{\cal A}}
\def\ket#1{\vert #1 \rangle}
\def\bra#1{\langle #1 \vert}

\def\ip#1#2{\langle #1 \vert #2 \rangle}
\def\me#1#2#3{\langle #1 \vert #2 \vert #3 \rangle}

\def\intbz{\int_{\rm BZ}\!\frac{d^3k}{(2\pi)^3}}
\renewcommand{\bf}{\mathbf}
\def\d{\partial}

\def\P{\mathcal{P}}
\def\Q{\mathcal{Q}}
\def\Tr{\mathrm{Tr}}
\def\tr{\mathrm{tr}}

\begin{document}
\title{Orbital magnetoelectric coupling in band 
insulators}
\author{Andrew M.~Essin}
\affiliation{Department of Physics, University of California,
Berkeley, CA 94720}
\author{Ari M.~Turner}
\affiliation{Department of Physics, University of California,
Berkeley, CA 94720}
\author{Joel E.~Moore}
\affiliation{Department of Physics, University of California,
Berkeley, CA 94720} \affiliation{Materials Sciences Division,
Lawrence Berkeley National Laboratory, Berkeley, CA 94720}
\author{David Vanderbilt}
\affiliation{Department of Physics and Astronomy, Rutgers University,
Piscataway, NJ 08854}
\date{\today}


%

\begin{abstract}
Magnetoelectric responses are a fundamental characteristic of
materials that
break time-reversal and inversion symmetries (notably
multiferroics) and, remarkably,
of ``topological insulators'' in which those
symmetries are
unbroken.  
Previous work has shown how to compute spin and lattice 
contributions to
the magnetoelectric tensor.  Here we solve the
problem of orbital contributions by
computing the frozen-lattice electronic polarization induced by a
magnetic field.
One part of this response (the ``Chern-Simons term'') can appear even in
time-reversal-symmetric materials
and has been previously shown to be quantized in topological insulators.
In general materials there are additional orbital contributions to all parts of the magnetoelectric tensor; these vanish in topological insulators by symmetry and also vanish in several simplified models  {\it without} time-reversal and inversion whose magnetoelectric couplings were studied before.
We give two derivations of the response formula, one based on a uniform magnetic field and one based on extrapolation of a long-wavelength magnetic field, and discuss some of the consequences of this formula.
\end{abstract}


\maketitle

\section{Introduction}

Understanding the response of a solid to applied magnetic or electric fields is
of both fundamental and applied interest.  Two standard examples are that 
metals can be distinguished from insulators by their screening of an applied 
electric field, and superconductors from metals by their exclusion of 
magnetic field (the Meissner effect).  Magnetoelectric response in insulators 
has been studied for many years and is currently undergoing a renaissance 
driven by the availability of new materials.  The linear response of this type 
is the magnetoelectric polarizability: in ``multiferroic'' materials that break
parity and time-reversal symmetries, an applied electric field creates a 
magnetic dipole moment and a magnetic field creates an electric dipole moment,
and several applications have been 
proposed for such responses.
Such responses are observed in a variety of
materials and from a variety of mechanisms.\cite{spaldin-s05,fiebig-jpdp05}
From a theoretical point of view, the most
intriguing part of the polarizability is that due to the
orbital motion of electrons, because the orbital motion couples
to the vector potential rather than the more tangible
magnetic field.

The orbital magnetoelectric polarizability has also been studied recently in
non-magnetic materials known as ``topological insulators.'' 
These insulators have Bloch wavefunctions with unusual 
topological properties, that lead to
a magnetoelectric response described by an ${\bf E} \cdot {\bf B}$ term in 
their effective electromagnetic Lagrangians,~\cite{qilong} 
with a quantized coefficient.  Qi, Hughes, and Zhang~\cite{qilong} (QHZ)
gave a formula for the coefficient of this term.  For the specific 
case of topological band insulators, their result reproduces earlier 
formulas for the relevant topological
invariant,~\cite{fu&kane&mele-2007,moore&balents-2006,rroy3d} but it
is more generally valid: it describes a contribution to the 
magnetoelectric polarizability non just in topological insulators but
in any band insulator.  Their formula has a periodicity or ambiguity
by $e^2/h$ that is related to the possibility of surface quantum Hall
layers on a three-dimensional sample and generalizes the ambiguity of
ordinary polarization.

The same ${\bf E} \cdot {\bf B}$ coupling, known as ``axion 
electrodynamics'' and originally studied in the 
1980s,~\cite{wilczekaxion} was obtained in a previous paper by three
of the present authors~\cite{essinomp} using a semiclassical 
approach~\cite{xiao} to compute $dP/dB$, the polarization response to
an applied magnetic field.  However, in a general material, that 
semiclassical approach leads to an explicit formula for only part of 
the orbital magnetoelectric polarizability, the part found by 
QHZ.~\cite{qilong}  The remainder, which is generically nonvanishing 
in materials that break inversion and time-reversal symmetries,
is expressed only implicitly in terms of the modification of
the Bloch wavefunctions by the magnetic field. 

In this paper, 
we develop a more microscopic approach that enables us to compute all terms in 
the orbital response explicitly in terms of the unperturbed wavefunctions,
thereby opening the door to realistic calculations using modern
band-structure methods (e.g., in the context of density-functional
theory).  Moreover, beyond
its importance for computation, this expression clarifies the physical
origins of the orbital magnetoelectric polarizability and resolves some issues
that arose in previous efforts to describe the ``toroidal moment'' in periodic
systems.

In the remainder of this introduction, 
we review some macroscopic features of the magnetoelectric response, while
subsequent sections will be devoted mainly to a detailed treatment of 
microscopic features.
The magnetoelectric tensor can be decomposed into trace and traceless
parts as
\begin{equation} \label{eq:alphadef}
\frac{\d P^i}{\d B^j} = \frac{\d M_j}{\d E_i} = \alpha^i_j 
= \tilde{\alpha}^i_j + \alpha_\theta \delta^i_j,
\end{equation}
where $\tilde{\alpha}$ is traceless and
\begin{equation}
\alpha_\theta = \frac{\theta}{2\pi} \frac{e^2}{h}
\end{equation}
is the trace part expressed in terms of the dimensionless parameter 
$\theta$; $\alpha$ has the physical dimension of conductance. The
trace is the most difficult term to determine, as its physical
effects are elusive. 
It should be noted that equality between $\d P^i/\d B^j$ and 
$\d M_j/\d E_i$ only holds in the absence of dissipation and 
dispersion, which describes the low frequency, low temperature 
responses of an insulator.~\cite{expttheta,hornreichshtrikman}
The placement of the indices in Eq.~(\ref{eq:alphadef})
is not essential for the arguments and calculations
in this paper, and the reader can choose to treat $\alpha$ as a Cartesian tensor
$\alpha_{ij}$ if desired.~\cite{foot1} As a Cartesian tensor,
the traceless part decomposes further into symmetric 
and antisymmetric parts
\begin{equation}
\tilde{\alpha}^S_{ij} = 
\frac{1}{2} \left( \tilde{\alpha}_{ij} + \tilde{\alpha}_{ji} \right), \quad
\tilde{\alpha}^A_{ij} = 
\frac{1}{2} \left( \tilde{\alpha}_{ij} - \tilde{\alpha}_{ji} \right)
=-\epsilon_{ijk} T_k,
\end{equation}
where $T_i = -\epsilon_{ijk} \tilde{\alpha}_{jk}/2$ is the toroidal response.
(Unless otherwise stated, in our work repeated indices are implicitly summed.)
The terminology reflects that this part of the orbital magnetoelectric response is related to the ``toroidal moment'', which is an order parameter that has recently been studied intensively; in a Landau effective free energy, the toroidal moment and the toroidal part of the magnetoelectric response are directly
related.~\cite{edererspaldin,aligia}

The primary goal of this paper is to compute the contribution to $\alpha$ that
arises solely from the motion of electrons due to their couplings to the
electromagnetic potentials $\rho\phi$ and $-\bf{j}\cdot\bf{A}$.  We call this
contribution the orbital magnetoelectric polarizability, or OMP for short.
Other effects, such as those mediated by 
lattice distortions or the Zeeman coupling to the electron's spin, are calculable 
with known methods.\cite{wojdel}  We shall only treat the
polarization response to an applied magnetic field here; concurrent work by
Malashevich, Souza, Coh, and one of us obtains an equivalent formula by
developing methods to compute the orbital magnetization induced by an
electrical field.\cite{malashevichOMP}

The magnetoelectric tensor's physical consequences arise through the bound 
current and charge,\cite{qilong,expttheta} given by 
$\rho_b=-\mathrm{div} \,\bf{P}$ and 
$\mathbf{J}_b=\d_t \bf{P} + \mathrm{curl} \,\M$.
Besides having a ground state value,
each moment responds 
(instantaneously and locally, as appropriate for the low-frequency
response of an insulator)
to applied electric and magnetic fields,
\textit{e.g.}, $P^i=P_0^i+\chi_E^{ij}E_j+\alpha^i_jB^j$; we will concentrate
on the magnetoelectric response. (Unless otherwise stated, in this article
repeated indices are implicitly summed.) It is useful
to allow $\alpha^i_j$, a material property, to vary in space and time by allowing
the electronic Hamiltonian to vary; this leads
to a formula that covers the effects of boundaries and time-dependent
shearing of the crystal, for example.  Then the relevant terms are
\begin{alignat}{2} \label{eq:boundcharge}
J^i_b 
&= (\tilde{\alpha}^l_j \epsilon^{ikj} - \tilde{\alpha}^i_j \epsilon^{jkl} ) 
\d_k E_l &&+ (\d_t \alpha^i_j) B^j + \epsilon^{ijk} (\d_j \alpha^l_k) E_l 
\notag \\
\rho_b &= - \tilde{\alpha}^i_j\d_i B^j &&- (\d_i \alpha^i_j)B^j,
\end{alignat}
We have used two of 
Maxwell's equations to simplify the first term in each line.
The most important point to notice here is that $\alpha_\theta$ does not appear
except in derivatives, so that any uniform and static contribution to $\theta$ has
no effect on electrodynamics. Hence
in a uniform, static crystal, the components of $\tilde{\alpha}$ can be 
computed or measured from the current or charge response to spatially varying
fields, given by the first term in each line.  On the other hand, if we wish 
similarly to obtain $\alpha_\theta$ from charge
or current responses to applied fields, we need to 
consider a crystal that varies either spatially or temporally, so that $\E$
or $\B$ will couple to $\d_i \alpha_\theta$ or $\d_t \alpha_\theta$, as in
the second terms of Eqs.~\eqref{eq:boundcharge}.  These considerations, which
we will elaborate later, motivate our theoretical approach to
the OMP in this paper.

We will proceed as follows.
In Section II, we present the results of our calculation of the OMP 
in the independent-electron approximation.
This section includes a review of known results, followed by
a discussion of the new contributions we compute and when those contributions
can be expected to vanish (so that the OMP reduces to the form found in
the literature previously).  We follow these discussions with a detailed
presentation of
the calculation in Section III.  This calculation involves a novel method for
dealing with a uniform magnetic field in a crystal. 
An alternative derivation is presented in the Appendix.

\section{General features of orbital magnetoelectric response}

In this section we discuss properties of the OMP and its 
explicit expression in the independent electron approximation.  There
is a natural decomposition into two parts, which is, however, not 
equivalent to the standard symmetry decomposition given in
Eq.~\eqref{eq:alphadef} of the Introduction.

The first part is the scalar 
``Chern-Simons'' term $\alpha_{CS}$ obtained by QHZ~\cite{qilong}
that contributes only to the trace part $\alpha_\theta$.  It is 
formally similar to the Berry-phase expression for 
polarization~\cite{ksv} in that it depends only on the wavefunctions,
not their energies, which explains the terminology ``magneto-electric polarization'' introduced by QHZ for $\alpha_{CS}$.~\cite{qilong}
The second part of the response is not simply scalar.  It 
has a different mathematical form that is not built from the Berry 
connection, looking like a more typical response function in that it 
involves cross-gap contributions and is not a ``moment'' determined 
by the unperturbed wavefunctions.  We label this term $\alpha_G$ 
because of its connection with cross-gap contributions.
This term does not seem to have been obtained previously although its physical
origin is not complicated.

\subsection{The OMP expression and the origin of the cross-gap term $\alpha_G$}

We first give the microscopic expression of the new term in the OMP
and discuss its interpretation.
The later parts of this section explain why the new term vanishes in 
most of the models that have been introduced in the literature to study 
the OMP, and discuss to what extent the two terms in the OMP expression 
are physically separate.
The OMP expression that we
discuss here will be derived later in Section III as follows:
we compute the bulk current in
the presence of a small, uniform magnetic field as the crystal Hamiltonian is 
varied adiabatically.  The result is a total time derivative which can be
integrated to obtain the magnetically induced bulk polarization.  

The most obvious property of the new term $\alpha_G$ in the response is that,
unlike the Chern-Simons piece, it has off-diagonal components;
for instance, $\d P^x/\d B^y\neq0$.
To motivate the expression for $\alpha_G$
intuitively, we note that it is very similar to what one would
expect based on simple response theory: An electric
dipole moment, $e\r$, is induced when a magnetic field is
applied. This field couples linearly to the magnetic dipole moment 
$(e/4)(\r\times\v-\v\times\r)$
(this form takes care of the operator ordering when we go to operators on Bloch states).
The expression we actually get for the OMP is
expressed in terms of the periodic part of the Bloch wave functions
$u_{l\k}$ and the energies $E_{l\k}$ describing the electronic structure of a 
crystal:
\begin{widetext}
\begin{subequations}
\begin{align}
\alpha^i_j &= (\alpha_G)^i_j + \alpha_{CS} \delta^i_j \\
(\alpha_G)^i_j &= 
\sum_{\begin{subarray}{c}n\,\mathrm{occ}\\m\,\mathrm{unocc}\end{subarray}}
\intbz \mathrm{Re} \left\{
\frac{ \me{u_{n\k}}{e \!\not\!r^i_{\k}}{u_{m\k}} 
\me{u_{m\k}}{
e(\mathbf{v}_{\k}\times\!\not\!\r_{\k})_j-e(\!\not\!\r_{\k}\times\mathbf{v}_{\k})_j 
-2i\d H'_{\k}/\d B^j 
}{u_{n\k}} 
}{E_{n\k}-E_{m\k}} 
\right\} \label{eq:alphaI} \\
\alpha_{CS} &= -\frac{e^2}{2\hslash} \, \epsilon_{abc}  \intbz \, \mathrm{tr} 
\left[ \mathcal{A}^a \partial^b \mathcal{A}^c 
- \frac{2i}{3} \mathcal{A}^a \mathcal{A}^b \mathcal{A}^c \right].
\label{eq:alphaCS}
\end{align} \label{eq:alphatot}
\end{subequations}
\end{widetext}
Here the Berry connection 
$\mathcal{A}^a_{nn'}(\k) = i \ip{u_{n\k}}{\d_{k_a}u_{n'\k}}$ is a matrix on the
space of occupied wave functions $u_{n\k}$, and the derivative with an upper index
$\d^a = \d_{k_a}$ is a $k$-derivative, as opposed to the spatial derivative 
$\d_i$ in Eq.~\eqref{eq:boundcharge}.  The velocity operator is related 
to the Bloch Hamiltonian, $\hslash v^i(\k)=\d^i H_{\k}$, while the 
operator $\!\not\!r^i_{\k}$ is defined as the derivative $\d^i\mathcal{P}_{\k}$
of the projection $\P$ onto the occupied bands at $\k$.
This operator is closely related to the position operator;
its ``cross-gap'' matrix elements between occupied and unoccupied bands
are $\me{u_{m\k}}{\negthickspace\not\negmedspace r^i_{\k}}{u_{n\k}} = 
\me{u_{n\k}}{\negthickspace\not\negmedspace r^i_{\k}}{u_{m\k}}^* = -i\me{u_{m\k}}{r^i}{u_{n\k}}$,
while its ``interior'' matrix elements between two occupied bands or two 
unoccupied bands 
vanish.  Finally, the operator $H'$ is introduced for generality, as
discussed in Section \ref{sec:mag}; it vanishes for the continuum 
Schr\"{o}dinger 
Hamiltonian and for tight-binding Hamiltonians whose hoppings are all 
rectilinear, and so will be ignored for most of the analysis that follows. 
Neglecting this subtlety, the form of $\alpha_G$ is nearly what would be 
expected for the response in electric dipole moment to a field coupling 
linearly to the magnetic dipole moment.  In the derivation 
presented in Section III, the term $\alpha_G$ appears in abbreviated
form at Eq.~\eqref{eq:alphaI2}, and $\alpha_{CS}$ follows immediately
from Eq.~\eqref{eq:JCSfinal}.

The main difference between the explicit form of $\alpha_G$ and the na\"ive
expectation from the dipole moment argument above is that $\alpha_G$ excludes 
terms of the form $\me{n}{\r}{m}\me{m}{\v}{n'}\me{n'}{\r}{n}$,
for example, that include interior matrix elements of $\r$.  In some
sense, this 
omission is compensated for by the extra factor of 2 
relative to the na\"ive expectation and by a
remainder term, namely,
$\alpha_{CS}$, the Chern-Simons part.
The Chern-Simons term $\alpha_{CS}$ alone has appeared
previously.\cite{qilong,essinomp}  The next subsection reviews the
properties of $\alpha_{CS}$ and gives a geometrical picture for its discrete
ambiguity, which is not present in the $\alpha_{G}$ term.  We then explain how 
the existence of the previously unreported $\alpha_G$ can be reconciled with
previous 
studies on model Hamiltonians that found only $\alpha_{CS}$, and then show that 
the two terms are more intimately related than they first appear.

\subsection{The Chern-Simons form, axion electrodynamics, and topological
insulators} \label{sec:CS}

The Chern-Simons response $\alpha_{CS}$ has been discussed at some length in 
the literature.\cite{qilong,essinomp}  It does not emerge as clearly as 
$\alpha_G$ from the intuitive argument above about dipole moment in a field;
rather, in Ref.~\onlinecite{essinomp}, it was derived by treating
the vector potential as a background inhomogeneity and utilizing a general 
formalism for computing the polarization in such a background.~\cite{xiao}

The most important feature of the microscopic expression for the 
\emph{isotropic} OMP is that
it suffers from a discrete ambiguity.  The dimensionless
parameter $\theta$ quantifying the isotropic susceptibility contains
the term
\beq
\theta_{CS} = -\frac{1}{4\pi} \, \epsilon_{abc}  \int \! d^3k \, \mathrm{tr} 
\left[ \mathcal{A}^a \partial^b \mathcal{A}^c 
- \frac{2i}{3} \mathcal{A}^a \mathcal{A}^b \mathcal{A}^c \right],
\eeq
which
is only defined up to integer multiples of $2\pi$.  This is tied to a ``gauge''
invariance: ground state properties of a band insulator should only be 
determined by the ground state density matrix $\rho^g_{\k}$, which is invariant 
under unitary transformations $U_{nn'}(\bf{k})$ that mix the occupied bands.
Now, the Berry connection $\A$ is not invariant under such a
transformation, but there is no inconsistency because,
in the expression for $\theta_{CS}$, all
the terms produced by the gauge transformation 
cancel except for a multiple of $2\pi$.
An analogous phenomenon, slightly easier to understand,
is found in the case of electric polarization\cite{ksv}
\beq
P^i = e\intbz\, \A^i \;,
\eeq
which has invariance only up to  a discrete ``quantum,'' or ambiguity, which 
counts the number of times $U(\bf{k})$ winds around the Brillouin zone
(e.g., if $U_{11}=e^{ik_x a}$ and $U_{ii}=1$, $i\neq 1$, then $P^x$ changes
by one quantum).   The 
Chern-Simons response $\alpha_{CS}$ behaves similarly, although the 
``winding'' that leads to the ambiguity is more complicated (in particular,
it is non-Abelian). 

These ambiguities can be understood from general arguments, without relying
on the explicit formulae.  
In the case of 
the polarization, the quantum of uncertainty of $P^x$,
$e/S_x$, depends on the lattice structure, with $S_x$ 
the area of a surface unit cell normal to $x$.  The ambiguity results
because the bulk polarization
does not completely determine 
the surface charge:
isolated surface bands can be filled or emptied, changing the number of
surface electrons per cell by an integer.
For the magnetoelectric response, the quantum of magnetoelectric polarizability
is connected with the fact that $\theta$ gives a surface Hall
conductance, as can be seen from the term $\bf{J}_b = (\bm{\nabla}
\alpha_\theta )\times \E$ in Eq.~\eqref{eq:boundcharge}.  Therefore, the
ambiguity in $\alpha_\theta$ is just $e^2/h$, the ``quantum of
Hall conductance,'' because it is
possible to add a quantum Hall layer to the surface.  (This
remains a theoretical possibility even if no intrinsic quantum Hall
materials have yet been found.)

Now let us show that this ambiguity afflicts only the trace of
the susceptibility. This can be seen directly by measuring the bound
charge and currents.  For example, 
all the components of $\tilde{\alpha}$ can be deduced from a measurement
of $\rho_b$ in the presence of a nonuniform magnetic field [see 
Eq.~(\ref{eq:boundcharge})], but $\alpha_\theta$ itself does not determine 
any bulk properties.

More concretely, one can 
derive the ambiguities in the magnetoelectric response from the ambiguities
in the surface polarization. 
In a periodic system, which for simplicity we take to have a cubic unit cell, 
the smallest magnetic field that can be applied without destroying
the periodicity of the Schr\"odinger equation corresponds to
one flux quantum per unit cell,
or $B = h / (e S)$, where $S$ is again a
transverse cell area.  The ambiguity in the polarization of the system in this
magnetic field corresponds to an ambiguity in $dP/dB$ of
\beq
{\Delta P \over B} = {e/S \over h/(eS)} = {e^2 \over h}.
\eeq
Hence on purely geometrical grounds there is
a natural quantum $e^2/h$ of the {\it diagonal} magnetoelectric
polarizability.~\cite{essinomp}

\begin{figure}
\includegraphics[width=0.8\columnwidth]{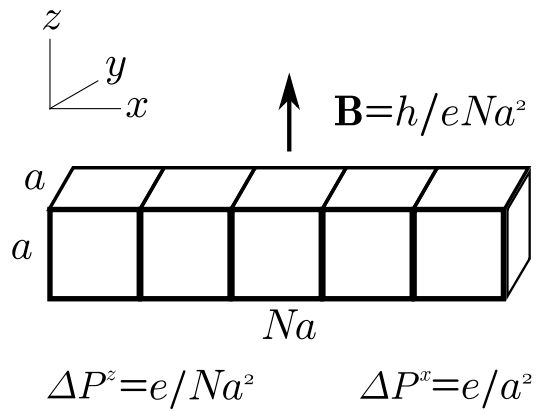}
\caption{A supercell admits a small magnetic flux, and the quantum of 
polarization transverse to the long direction is correspondingly small, but
the quantum for polarization along the long direction is much larger.
\label{polargeomfig}}
\end{figure}

In order to see that this uncertainty remains the same when
a \emph{small} magnetic field is applied (after all, $\alpha$ is defined
as a \emph{linear} response), we will have to construct
large supercells in a direction perpendicular to the applied $B$
(Fig.~\ref{polargeomfig}).
While a supercell of $N$ fundamental cells has a 
less precisely defined polarization
(the quantum decreases by a factor
$N$, so the \emph{uncertainty} increases), the minimum field that can be applied also decreases by this factor, so 
that the uncertainty in the polarizability $dP^i/dB^i$ (no sum) remains 
constant.  On 
the other hand, if we consider the off-diagonal response, we can consider a
supercell with its long axis parallel to the applied $B$.
In this case, the polarization quantum remains constant as the supercell grows
large and the minimum applied flux becomes small; the quantum in $dP^i/dB^j$ 
(for $i \not = j$) then becomes large, which means that the uncertainty
vanishes.  For this geometry, a small $B$ acts like a continuous parameter,
and the change in polarization induced by $B$ can be continuously tracked, even
if the absolute polarization remains ambiguous.

Thus, with or without interactions, there is a fundamental
difference between the isotropic response and the other components
of the response.
For the trace-free components, we indeed do not find a quantum of
uncertainty in the polarizability formula.  In particular, if  
the toroidal response is defined by
 $T_i = -\epsilon_{ijk} \tilde{\alpha}_{jk}/2$, then we 
believe that a
``quantum of toroidal moment''~\cite{aligia} can only exist
when there is a spin direction with conserved ``up'' and ``down''
densities.  (This toroidal moment is typically defined as 
$\bm{t} = (1/2)\int \r\times\bm{\mu}d\r$, with $\bm{\mu}$ the 
magnetization density,\cite{edererspaldin} or more generally in terms of a 
tensor $\mathcal{T}^{ij}$ such that 
$\d_i\mathcal{T}^{ij}=-2\mu^j$.\cite{aligia})
It then reduces to the polarization difference between up 
and down electrons.

A particular class of materials for which the
ambiguity in $\alpha_{\theta}$ is extremely important
is the strong topological 
insulators,\cite{fu&kane&mele-2007,moore&balents-2006,rroy3d} in which
$\theta = \pi$ (Ref.~\onlinecite{qilong}).
These are time-reversal ($T$) symmetric band
insulators.  At first blush, $T$ invariance should rule out magnetoelectric
phenomena at linear order, since $\M$ and $\B$ are $T$-odd.  However, the 
ambiguity by $2\pi$ in $\theta$ provides a loophole, since $-\pi$ is equivalent
to $\pi$.  
Here we regard the ambiguity/periodicity of $\theta$ as a consequence of its microscopic origin (alternately, its coupling to electrons); because $\theta$ can be modified by $2 \pi n$ by addition of surface integer quantum Hall layers, only $\theta$ modulo $2 \pi$ is a meaningful bulk quantity for systems with charge-$e$ excitations.  This is consistent with the gauge-dependence of the integral for $\alpha_{CS}$.  An alternate approach is to derive an ambiguity in $\theta$ by assuming that the $U(1)$ fields are  derived from a non-Abelian gauge field.\cite{wilczekaxion}
The view here that periodicity of $\theta$ results from the microscopic coupling to electrons is similar to the conventional understanding of the polarization quantum.

\subsection{Conditions causing $\alpha_G$ to vanish}

It is worthwhile to understand in more detail the conditions
under which the response $\alpha_G$ is allowed.  It is forbidden in systems with inversion ($P$) or time-reversal ($T$) symmetry, which can be seen explicitly from the presence
of three $k$-derivatives acting on gauge-invariant matrices
in the formula written in terms of $\P_{\k}$ and 
$H_{\k}$.~\cite{foot2}
However, this alone is not sufficient to explain why $\alpha_G$ did not appear 
in the $T$-breaking models 
previously introduced to study the OMP.\cite{essinomp,qilong,dynamicalaxion} 
This is explained by the fact that
the interband contribution $\alpha_G$ [Eq.~(\ref{eq:alphaI})] will also vanish
if dispersions satisfy the following ``degeneracy" and ``reflection" 
conditions:
\begin{itemize}
\item At a given $\bf k$, all the occupied valence bands have the
same energy $E^{\rm v}_{\bf{k}}$.
\item Similarly, all the unoccupied conduction bands have the
same energy $E^{\rm c}_{\bf{k}}$.
\item $E^{\rm v}_{\bf{k}}+E^{\rm c}_{\bf{k}}$ is independent of $\bf k$
(and can be taken to be zero).
\end{itemize}
This can be seen immediately in an expanded form of the integrand of 
$\alpha_G$, [see Eqs.~\eqref{eq:unoccb} and \eqref{eq:unoccc}]
\begin{multline}
-\sum_{\begin{subarray}{c}n,n'\,\mathrm{occ}\\m\,\mathrm{unocc}\end{subarray}}
(E_n-E_{n'})\frac{\ip{\d^b n}{n'}\ip{\d^a n'}{m} \ip{m}{\d^i n} }{E_n-E_m} \\
+\sum_{\begin{subarray}{c}n\,\mathrm{occ}\\m,m'\,\mathrm{unocc}\end{subarray}}
(E_m-E_{m'})\frac{\ip{\d^b n}{m'}\ip{m'}{\d^a m} \ip{m}{\d^i n} }{E_n-E_m} \\
-\sum_{\begin{subarray}{c}n\,\mathrm{occ}\\m\,\mathrm{unocc}\end{subarray}}
\d^b(E_n+E_m)\frac{\ip{\d^a n}{m} \ip{m}{\d^i n} }{E_n-E_m},
\end{multline}
where $\ket{n} = \ket{u_{n\k}}$ and $E_n = E_{n\k}$, etc.
Such a structure is automatic when only two orbitals (with
both spin states) are taken into account and
the system has particle-hole and $PT$ symmetries.
$PT$ symmetry guarantees
that the bands remain spin-degenerate even if spin is not a good quantum
number.  To see this, recall that $T$ acts on wave functions as $i\sigma^y K$
and maps $\bf{k} \rightarrow -\bf{k}$.  Here, $K$ is complex conjugation and
$\sigma^y$ takes the form of the usual Pauli matrix in the $z$ basis of spin.
Then $P$ maps $\bf{k} \rightarrow -\bf{k}$ again, so that $PT$ effectively acts
as ``$T$ at each $\k$.''\cite{asss}  
Then particle-hole symmetry implies that the dispersion is
reflection-symmetric, $E^{\rm v}_{\bf{k}}=-E^{\rm c}_{\bf{k}}$.

\begin{figure}
\includegraphics[width=0.8\columnwidth]{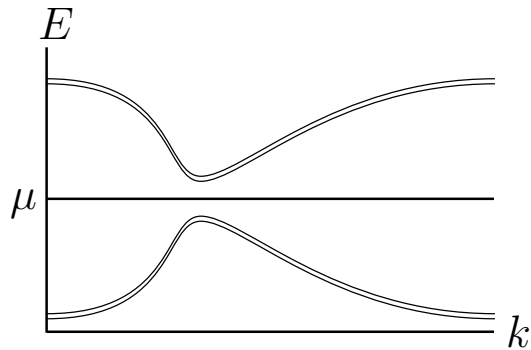}
\caption{Schematic band structure that leads to vanishing $\alpha_G$.  The 
bands below the chemical potential are degenerate with energy 
$E^{\rm v}_{\bf{k}}$, while 
the bands above the chemical potential have energy 
$E^{\rm c}_{\bf{k}} = \mathrm{const}. - E^{\rm v}_{\bf{k}}$.
\label{fig:symspec} }
\end{figure}

Most model Hamiltonians discussed in the literature that access the
topological insulator 
phase,\cite{fu&kane&mele-2007,qilong,essinomp,hosur,dynamicalaxion}
as well as the Dirac Hamiltonian (in the context of which the axion electrodynamics
was first discussed\cite{wilczekaxion}),
can be defined in terms of a Clifford algebra,~\cite{foot3}
and this ensures that the dispersions are degenerate and
reflection symmetric.
The only exception of which we are aware is the model of Guo and Franz
on the pyrochlore lattice, which has four orbitals per unit cell.\cite{guo}  
The topological insulator
phase itself will not have a contribution from $\alpha_G$, since it is
$T$-invariant, and so the Guo and Franz model will not show such a response;
however, the addition of any $T$-breaking perturbation to their model
should produce off-diagonal magnetoelectric responses.

Finally, there is a simple mathematical condition that will cause $\alpha_G$ 
to vanish.  Namely, $\alpha_G$ decreases as the gap becomes large without
changing the wave functions,
and in the limit of infinite bulk gap the only magnetoelectric response comes 
from the Chern-Simons part, which is not sensitive to the energies and depends
only on the electron wave functions.

\subsection{Is the Chern-Simons contribution physically distinct?}
Apart from the ambiguity in $\alpha_{CS}$ that is not present in 
$\alpha_G$, there seems to be no real physical
distinction between the two terms of the linear magnetoelectric response.
We discuss two aspects that relate to this observation below.

\emph{Localized vs. itinerant contributions}

The ambiguity in $\theta_{CS}$ can be interpreted as a manifestation of
the fact that bulk quantities cannot determine the surface quantum Hall 
conductance, since a two-dimensional quantum Hall layer could appear on a 
surface independent of bulk properties.  This suggests, perhaps, that the
Chern-Simons term appears only in bulk systems with extended wave functions,
and is a consequence of the itinerant electrons,
while $\alpha_G$ is a localized molecule-like contribution.
However, this turns out not to be the case.

\begin{figure}
\includegraphics[width=0.6\columnwidth]{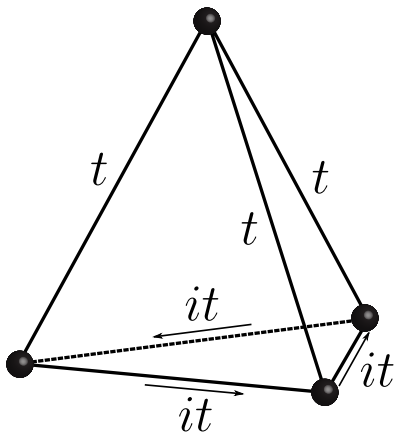}
\caption{A tetrahedral tight-binding molecule for spinless electrons, with one
orbital per site and 
complex hoppings.  The hopping integrals are all equal, except that those 
around one face have a phase of $i$ relative to the other three.  There are
then two pairs of degenerate levels.
\label{fig:molecule}}
\end{figure}

Consider a periodic array of isolated molecules, which is an extreme limit of
the class of crystalline insulators.  Such a system has flat bands, with
energies equal to the energies of the molecular states, since the
electrons cannot propagate.  It is certainly possible to construct a molecular
system where all the unoccupied states have
one energy and all the occupied states have another, by tuning the potentials. 
In this case $\alpha_G$
will vanish.  However, such a molecule can still display a magnetoelectric
response; it will therefore have to be given by $\alpha_{CS}$ (and so
restricted to diagonal responses).  For example, consider the ``molecule'' 
of Fig.~(\ref{fig:molecule}) with the shape of a regular tetrahedron.
If the two low-energy levels are occupied, the magnetoelectric response is
\begin{equation*}
\frac{\d P^i}{\d B^j} = \pm \delta^i_j \frac{1}{\sqrt{6}} \frac{e^2}{\hslash},
\end{equation*}
where $P^i$ here is the electric dipole moment divided by the volume of the 
tetrahedron; the sign of the polarizability reverses when the complex phases
are reversed.  This shows
that the Chern-Simons term does not require delocalized orbitals.

\emph{Additivity}

Another argument against distinguishing
between the Chern-Simons part and the rest of the susceptibility
is based on band additivity.
When interactions are not taken into account, each occupied band can
be regarded as an independent physical system
(at least if
there are no band crossings).  Applying a magnetic
field causes each band $n$ to become polarized by a certain amount $\bf{P}_n$,
and so the net polarization should be $\bf{P}=\sum_n \bf{P}_n$. The
Pauli exclusion principle does not lead to any ``interactions"
between pairs of bands, because the polarization (like any single-body
operator) can be written as the sum of the mean polarization in
each of the orthonormal occupied states.

Now the Chern-Simons form does not look particularly
additive in this sense, and is not by itself.
Because it is the trace of a
matrix product in the occupied subspace, it necessarily involves matrix
elements between different occupied states, while an additive formula would
not. Nevertheless, $\alpha_G$ and $\alpha_{CS}$ are together additive, as can
be seen most simply in Eq.~\eqref{eq:compactalpha}, where the two terms combine
into a single sum over occupied bands.
In terms of $\alpha_G$ and $\alpha_{CS}$ separately, one finds that when the
values of $\alpha_G$, assuming just band
1 or 2 is occupied, are added together, some terms occur that
are not present in the expression for $\alpha_G(1+2)$ (where
both bands are occupied), and vice-versa. 
Using Eqs.~\eqref{eq:unoccb} and \eqref{eq:unoccc}
we see, in fact, that
$\alpha_G$ is a sum of contributions which depend on three bands, as
$\alpha_G=\sum_{n,m,m'} C(n;m,m') + \sum_{n,n',m} D(n,n';m)$.  Terms such
as $C(1;2,m')$ are not
present in the expression for $\alpha_G(1+2)$. (Likewise
$D(1,2;m)$ appears in $\alpha_G(1+2)$ but not in
$\alpha_G(1)$ and $\alpha_G(2)$.)   Adding up the
discrepancies, 
one finds that the energy-denominators all cancel,
and the non-diagonal terms from the Chern-Simons form appear!

Seemingly paradoxical is the fact that for band structures satisfying
the degeneracy and reflection conditions of the last subsection, 
the magnetoelectric susceptibility
is given by the Chern-Simons term alone, which does not seem to be additive. 
However, the additivity property applies only to bands that do not cross.
It does not make any sense to ask whether the susceptibility is the sum over 
the susceptibilities for the systems in which just one of the degenerate bands 
is occupied, since those systems are not gapped.

\section{The OMP as currents in response to Chemical Changes}

Now we will tackle the problem of deriving the formula for
the OMP $\alpha$ discussed in the
last section.  There are two impediments we need to overcome, a physical
one, and a more technical one (which we will overcome starting
from an insight of
Levinson).\cite{levinson}

In order to determine $\alpha$, we would like to carry out a thought
experiment in which a crystal is exposed to appropriate electromagnetic
fields. For specificity, we will apply a uniform magnetic field.
To make the calculation of the response
clean, we wish to deal with an infinite crystal.
Then the polarization does not simply reduce to the first moment of the 
charge density,\cite{ksv}
so we will instead have to calculate the current or
charge distribution induced by the fields, and then use 
Eq.~(\ref{eq:boundcharge}) to deduce $\alpha$. If both the crystal
and the electromagnetic fields are independent of space and time,
there is no macroscopic charge
or current density. We will assume \emph{spatial} uniformity,
so that there are two choices for how to proceed. Either the magnetic
field can be varied in time or the crystal parameters, and thus
$\alpha$, can be varied. In either case,
we measure the current that flows through the bulk and try
to determine $\alpha$.
As ever, the diagonal response $\alpha_\theta$ is the 
most difficult to capture: while either time-dependent
experiment can be used to determine $\tilde{\alpha}$, only
the latter approach sheds light on the value of
$\alpha_\theta$.  

\begin{figure}
\includegraphics[width=0.85\columnwidth]{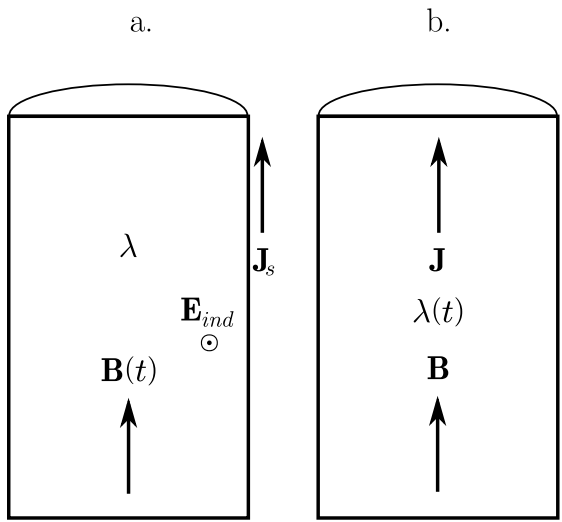}
\caption{As outlined in the text, (a) turning on a magnetic field produces
a macroscopic polarization through the flow of surface currents, while (b)
varying the crystal Hamiltonian in the presence of a fixed magnetic field
produces a polarization through the flow of current through the bulk.
\label{fig:currents}}
\end{figure}

To see why $\alpha_\theta$ can be determined only in this
way (given that we want to work with a spatially homogeneous geometry), 
let us discuss how currents flow through the
crystal.  The necessity of varying the crystal
in time can be deduced
from Maxwell's equations (see below) but we will give a more intuitive
discussion here. Suppose that $\tilde{\alpha}=0$.
Then in an applied magnetic field there is a polarization 
$\bf{P}=\alpha_\theta \B$; thus the crystal
gets charged at the surface.  As the magnetic field is turned on,
this surface charge has to build up (charge density 
$\hat{\bf{n}}\cdot\bf{P}$).
This
occurs entirely due to flows of charge {\it along the surface}.
Suppose, for example, that the sample is a cylinder (radius $R$) with the
magnetic field along its $z$ axis, as illustrated in
Fig.~\ref{fig:currents}(a).
Then an
electric field $\E_{\mathrm{ind}}=-\dot{B}R\hat{\bm{\phi}}/2$ is induced at
the surface according to Faraday's law.  Besides being the magnetoelectric
response, $\theta$ also represents the Hall coefficient for surface currents.
Therefore, a current of $J_s=\alpha_\theta\dot{B}R/2$ flows to the
top of the cylinder, adding up to a surface charge of
$2\pi R\int \!J_s(t)dt=\alpha_\theta B_f\pi R^2$ and producing the entire
polarization
$\alpha_\theta B_f$.  No current flows through the bulk!
In fact, the Hall conductance on the circular face 
produces a radial current as well, so that the charge distributes over the 
surface rather than just accumulating in a ring.
Note that the surface current
grows with the radius of the cylinder. This sounds like a nonlocal
response, but it can be understood as follows:
the electric field is determined by the non-local
Faraday law, but the \emph{crystal's} response to the electric field (namely,
the surface current) is local.

The current distribution can be understood directly from
Maxwell's equations: there are two contributions to the bulk
current, $\d_t \mathbf{P}$ and $\mathrm{curl}\ \mathbf{M}$.
The polarization is $\alpha_\theta \mathbf{B}$ while the
magnetization is indirectly produced by the induced
electric field, $\alpha_\theta \mathbf{E}_{\mathrm{ind}}$. The two contributions
thus cancel by Faraday's law in the bulk: 
$\mathbf{J}_{b}^{\mathrm{bulk}}
=\alpha_\theta(\d_t \mathbf{B}+\mathrm{curl}\ \mathbf{E})=0$.  
There \emph{is} a surface
current because 
$\alpha_\theta$ is discontinuous there.

On the other hand, if $\theta$ changes in time, while
the magnetic field is time-independent (as in Fig.~\ref{fig:currents}b), 
the polarization at the
ends of the cylinder builds up entirely by means of flows
of charge \emph{through the bulk}. Surface
flows cannot be large enough to explain the net polarization in
this situation. Since there is no induced electric field, the
surface current is just proportional to the lateral surface area and
is negligible compared to the bulk
current. Therefore the bulk current is equal to
$(\d_t\alpha_\theta) B$ and can be integrated
to give $\alpha_\theta B$.

For the other component of the OMP, $\tilde{\alpha}$,
either thought experiment can be used.
The simplest approach, however, is still the crystal-variation method,
since the surface currents
are negligible in that case,~\cite{foot4} and in
any case this method allows us to find all the components
of $\alpha$ simultaneously.

\emph{Difficulties with the operator $\r$ and uniform magnetic fields}

There are two technical difficulties in the theory.
First, the operator $\r$ has unbounded
matrix elements and thus the matrix elements of
the magnetic dipole
moment $(e/4)(\mathbf{v}\times\r-\r\times\mathbf{v})$ are not
well-defined. This rules out the straightforward use of perturbation
theory to calculate
the electric dipole moment of an infinite crystal
in a uniform magnetic field. Second, if we consider
a crystal in a uniform magnetic field, Bloch's
theorem does not hold. Although
the magnetic field is uniform, the vector potential that appears in
the Hamiltonian depends on $\r$.

We avoid the problems of $\r$ as follows.  
The key idea is to work with the ground state density
matrix, rather than wave functions.  The individual eigenstates change
drastically when a magnetic field, no matter how small,
is applied (consider the difference
between a plane wave and a localized Landau level).  However, the
\emph{density matrix} of an insulator
summed over the occupied bands only changes by a small
amount when $\mathbf{B}$ is applied; over short distances the magnetic
field cannot have a strong effect (even in the example of Landau levels),
and the density matrix has only short-range correlations because
it describes an insulating state.
More technically, we show (subsection \ref{sec:mag})
that the broken translation invariance 
of any single-body operator $\mathcal{O}$ (such as the density
matrix) can be dealt with
by factoring out an Aharonov-Bohm-like phase from
its matrix $\mathcal{O}_{\r_1\r_2}$. This solves the problem of the
nonuniform gauge field and
leads to
expressions that depend only on differences between $\mathbf{r}$'s.
In addition, since the exponentially decaying
ground-state density matrix appears multiplying every expression,
the factors of $\mathbf{r}_1-\mathbf{r}_2$
are suppressed.

The calculation then proceeds as follows.  First, using the
symmetries of the electron Hamiltonian in a uniform magnetic field, we find
how the density matrix changes in a weak magnetic field.  
Next we compute the current response to an adiabatic variation of the
crystal Hamiltonian.  Finally, we show that this current can be expressed as 
a total time derivative, and therefore can be integrated to give the 
polarization; at linear order in $\B$ we can read off the coefficients,
the magnetoelectric tensor $\alpha$.

\subsection{Single-body operators for a uniform magnetic field} \label{sec:mag}
Recall the form of the Schr\"{o}dinger Hamiltonian for a single electron in
a crystal and under the influence of a magnetic field,
\beq
\label{eq:eAoverc}
H_S(\bf{p},\r) = \frac{1}{2m} \left[ \bf{p} - e \bf{A}(\r) \right]^2 + V(\r),
\eeq
where $V(\r+\bf{R}) = V(\r)$ for lattice vectors $\bf{R}$.  The necessity
of using the
vector potential $\bf{A}$ seems at first to spoil 
the lattice translation symmetry one would expect in a uniform
magnetic field.  However, as
noted by Brown\cite{brown} and Zak,\cite{zak} a more subtle form of translation
symmetry remains.  In particular, choosing the gauge 
\beq
\bf{A} = \frac{1}{2} \B \times \r,
\eeq
the Hamiltonian has ``magnetic translation symmetry'':
\beq
H_S(\bf{p},\r+\bf{R}) 
= e^{ie\B\cdot(\bf{R}\times\r)/2\hslash} H_S(\bf{p},\r) 
e^{-ie\B\cdot(\bf{R}\times\r)/2\hslash}.
\eeq

This condition defines magnetic translation symmetry for general single-body
operators.
Any operator $\mathcal{O}$ possessing this symmetry can be written 
in the position basis as
\begin{subequations}
\beq
\mathcal{O}_{\r_1\r_2} = \bar{\mathcal{O}}_{\r_1\r_2} 
e^{-ie \B\cdot(\r_1\times\r_2)/2\hslash },
\eeq
where $\bar{\mathcal{O}}$ has lattice translation invariance, 
\beq
\bar{\mathcal{O}}_{\r_1+\bf{R},\r_2+\bf{R}} = \bar{\mathcal{O}}_{\r_1\r_2}.
\eeq
\end{subequations}
Note that the phase is just $(ie/\hslash)\int d\bm{\ell}\cdot\bf{A} $ calculated
along the straight line from $\r_2$ to $\r_1$, which agrees with the intuition
that comes from writing the second-quantized form of the operator,
\beq \label{eq:straightline}
\mathcal{O} = \int\! d^3r_1 d^3r_2 \, \bar{\mathcal{O}}_{\r_1\r_2} 
c^\dag_{\r_1}e^{-ie \B\cdot(\r_1\times\r_2)/2\hslash }c_{\r_2}.
\eeq

This argument shows how to couple general Hamiltonians to uniform fields:
$H=\exp[(ie/\hbar)\int_{\mathbf{r}_2}^{\mathbf{r}_1} 
d\bm{\ell}\cdot\mathbf{A}]
[H_{0\mathbf{r}_1\mathbf{r}_2}+H'_{\mathbf{r}_1\mathbf{r}_2}(B)]$. 
The vector potential appears explicitly only in $\mathbf{A}$,
while $H'(B)$ gives the rest of the dependence on the magnetic field.
The Schr\"{o}dinger Hamiltonian (\ref{eq:eAoverc})
is obtained if we take
\beq
\bar{H}_{0,\r_1\r_2} = \left[ -\frac{\hslash^2}{2m}\nabla^2_{\r_2} 
+ V(\r_2) \right] \delta^{(3)}(\r_2-\r_1).
\eeq
and set $H'=0$. Our results also apply to tight-binding models.
We introduce
$H'$ to capture the possibility that in a tight-binding model the hoppings will
not be rectilinear, and hence that the phases in Eq.~\eqref{eq:straightline}
do not capture the full field dependence of the Hamiltonian.

\subsection{The ground state density operator}

We find it convenient to work with the one-body density matrix $\rho^g$, or
equivalently the projector onto the occupied states, whenever possible, 
because it is a
basis-independent object.  Also, in an insulator, $\rho^g_{\r_1\r_2}$ is
exponentially suppressed in the distance $|\r_2-\r_1|$, which tempers the
divergences that arise from the unboundedness of $\r$.\cite{kohnloc}
In any case, if the ground state is translationally symmetric, 
the structure described
above will apply to $\rho_g$ and we can be sure that \emph{the density
matrix has translational symmetry apart from a phase}:
\beq
\rho^g_{\r_1\r_2} = \bar{\rho}^g_{\r_1\r_2}
e^{-ie \B\cdot(\r_1\times\r_2)/2\hslash },\label{eq:ahronov}
\eeq
where $\bar{\rho}^g$ possesses the translation symmetry of the crystal lattice
and hence should connect smoothly to the field-free density matrix.  
Hence we will
write
\beq
\bar{\rho}^g = \rho_0 + \rho',
\eeq
where $\rho_0$ is the density operator of the crystal in the absence 
of the 
magnetic field.

\emph{Density matrix perturbation theory:}
Now we have to calculate $\rho'$, using a kind of
perturbation theory that focuses on density matrices rather
than wave functions, since the wave-functions suffer from the
problems discussed above. This perturbation theory
starts from two characteristic properties of the
density matrix: it commutes with $H$, and for fermions at zero
temperature it is a projection operator.
The latter means that 
all states are either occupied or unoccupied, so the eigenvalues of 
the density operator are 0 and 1, which is formalized as
\beq
\rho^g\rho^g = \rho^g
\eeq
(idempotency).\cite{diracdensity}  Expressed in the position basis,
\begin{align}
\rho^g_{\r_1\r_3} &= \int d\r_2 \rho^g_{\r_1\r_2}\rho^g_{\r_2\r_3} \notag \\
\bar{\rho}^g_{\r_1\r_3}
&= \int d\r_2 \bar{\rho}^g_{\r_1\r_2}\bar{\rho}^g_{\r_2\r_3} 
e^{-(ie/2\hslash)\B\cdot(\r_1\times\r_2+\r_2\times\r_3+\r_3\times\r_1)}.
\end{align}
The exponent is just $-i \phi_{123}/\phi_0$, proportional to
the magnetic flux through triangle
123, and the exponential can be expanded for small $B$.  At first order this 
gives
\begin{multline} \label{eq:rhoprime}
\rho'_{\r_1\r_3} = \int d\r_2 [\rho'_{\r_1\r_2} \rho_{0\r_2\r_3} 
+ \rho_{0\r_1\r_2} \rho'_{\r_2\r_3} \\
- \rho_{0\r_1\r_2} \rho_{0\r_2\r_3} \left(i\phi_{123}/\phi_0\right) ].
\end{multline}
\emph{The problem of the unbounded }$\r$'s \emph{is resolved} in 
this equation because the area $A_{123}$ of the triangle is finite and 
independent of the origin, and also suppressed by the factor of $\rho$. 

\emph{Calculation of} $\rho'$:
In the last term of Eq.~\eqref{eq:rhoprime}, we can rewrite 
$2A_{123}=\r_1\times\r_2+\r_2\times\r_3+\r_3\times\r_1 = (\r_2-\r_1)\times(\r_3-\r_2)$
and then use $(\r_2-\r_1)\rho_{0\r_1\r_2} = [\rho_0,\r]_{\r_1\r_2}$, etc., to
obtain 
\beq \label{eq:exclusion}
(1-\rho_0)\rho'(1-\rho_0) - \rho_0\rho'\rho_0 
= -i\frac{e}{2\hslash}\B\cdot([\rho_0,\r]\times[\rho_0,\r]).
\eeq
If we define
\beq
\bar{H} = H_0 + H',
\eeq
then analogous manipulations (including 
$(\r_1-\r_2)H_{\r_1\r_2} = i\hslash \bf{v}_{\r_1\r_2}$) on the equation 
$[H,\rho^g] = 0$ give
\beq
[\rho',H_0] = \frac{e}{2}\mathbf{B}\cdot 
([\rho_0,\r]\times\mathbf{v}-\mathbf{v}\times[\rho_0,\r]) - [\rho_0,H'].
\label{eq:Lorentz}
\eeq

Eqs.~(\ref{eq:exclusion}) and (\ref{eq:Lorentz}) have an
intuitive meaning. The former equation determines the ``interior'' matrix
elements of $\rho'$, those between two occupied or two unoccupied
states of the zero-field Hamiltonian.
An perturbation with the full crystal symmetry does not change the interior 
matrix elements of the density matrix because
of the exclusion principle.\cite{mcweeny}  In our case, however, multiplying
$\rho_0$ by the phase $e^{(ie/2\hslash)\B\cdot\r_1\times\r_2}$ gives a density
matrix
with the correct magnetic translation symmetry, but also changes the momentum 
of the states and so results in a small probability for states to be douby 
occupied.  Therefore $\rho'$ must
correct for this ``violation of the exclusion 
principle.''  On the other hand Eq.~(\ref{eq:Lorentz}) determines
the ``cross-gap'' matrix elements of $\rho'$ (those between unoccupied
and occupied states).  These matrix elements capture the
expected ``transitions across the gap'' induced by the field.
The rest of this section is devoted to calculating all these matrix elements.
The results are given in Eqs.~\eqref{eq:Xanswer} and \eqref{eq:Lanswer}; the
derivations could be skipped on a first reading.

\emph{Calculation of\ }$\rho'$.
Precisely speaking, Eq.~(\ref{eq:exclusion})
gives the matrix elements of $\bar{\rho}^g$ 
between pairs of occupied ($n$ and $n'$) or unoccupied ($m$ and $m'$) states:
\begin{align} \label{eq:Xanswer}
\me{\psi_{n\k}}{\bar{\rho}^g}{\psi_{n'\k}} &= \delta_{nn'}
-\frac{e}{4\hslash} B^j \epsilon_{jab} \F^{ab}_{nn'} (\k) \notag\\
\me{\psi_{m\k}}{(1-\bar{\rho}^g)}{\psi_{m'\k}} &= \delta_{mm'}
-\frac{e}{4\hslash} B^j \epsilon_{jab} \check{\F}^{ab}_{mm'} (\k),
\end{align}
where $\F$ is the non-Abelian Berry curvature associated with the occupied 
bands,
\begin{align}
\F^{ab}_{nn'} 
&= i \me{u_{n\k}}{\d^a\P_\k\d^b\P_\k-\d^b\P_\k\d^a\P_\k}{u_{n'\k}} \notag \\
&= \d^a \A^b_{nn'} - \d^b \A^a_{nn'} -i [\A^a,\A^b]_{nn'},
\end{align}
and $\check{\F}$ is the corresponding quantity for the unoccupied bands.
To derive these relations, we use
\beq
\rho^g  = \intbz e^{i\k\cdot\r} \P_{\k} e^{-i\k\cdot\r}
\eeq
where $\P = \sum_{n\,\mathrm{occ}} \ket{u_{n\k}}\bra{u_{n\k}}$ is the 
projector onto filled bands at $\k$.  This gives
\begin{align} \label{eq:notr}
i[\rho^g,\r] 
&= \intbz e^{i\k\cdot\r} (\bm{\nabla}_{\k}\P_{\k} ) e^{-i\k\cdot\r}
\notag \\ &= \intbz e^{i\k\cdot\r} \not\r_\k e^{-i\k\cdot\r}
\end{align}
after discarding a total derivative.  The notation 
$\not\!\!\r = \bm{\nabla}_{\k}\P$ was introduced in Eq.~\eqref{eq:alphatot}.

By contrast, Eq.~\eqref{eq:Lorentz} 
describes to what extent
$\bar{\rho}^g$ fails to commute with $H_0$, the crystal Hamiltonian, and
gives the matrix elements of $\rho'$ 
between occupied and unoccupied states.  In this sense it is analogous to the 
more usual results for density-matrix perturbation theory.\cite{mcweeny}
In the basis of unperturbed energy eigenstates, 
\begin{multline} \label{eq:Lanswer}
\me{\psi_{n\k}}{\rho'}{\psi_{m\k}} = i\frac{e}{2\hslash} B^j \epsilon_{jab} 
\frac{ \me{u_{n\k}}{\{\d^a \P_{\k},\d^b H_{\k} \}}{u_{m\k}} }{E_{n\k}-E_{m\k}}
\\
+ \frac{ \me{u_{n\k}}{H'_{\k}}{u_{m\k}} }{E_{n\k}-E_{m\k}}.
\end{multline}
Recall that $\hslash v^b = \d^b H_{\k}$ and that $H'$ is introduced 
only to capture unusual situations such as 
tight-binding models with non-straight hoppings, and vanishes for the 
continuum problem.
Eqs.~\eqref{eq:Xanswer} and \eqref{eq:Lanswer} are the key technical results 
of this formalism, good to linear order in the magnetic field.  

\subsection{Adiabatic current}

Now we need to calculate the current as the Hamiltonian is changed 
slowly as a function of time, as in the ordinary theory of
polarization.\cite{resta,ksv}  
We have to be careful, however, since the current vanishes in the
zero-order adiabatic ground state described by density matrix
$\rho^g(t)$.  It is
necessary to go to first order in adiabatic perturbation theory, which
takes account of the fact that the true dynamical density matrix $\rho(t)$
has an extra contribution proportional to $dH/dt=\dot{H}$.
However, once the current has been expressed in terms of
$\dot{\rho}$, which is proportional to $\dot{H}$, the distinction
is no longer important and the adiabatic approximation can be made.

\emph{Preparing for the adiabatic approximation:}
We can write the current as
\begin{equation}
\mathbf{J}(t) = \frac{e}{\Omega} \Tr \rho(t)\mathbf{v} 
= \frac{e}{\Omega} \frac{i}{\hslash}\Tr \rho[H,\r]
\label{eq:comm1}
\end{equation}
where $\Omega$ is the crystal volume.
Here $H$ is the full Hamiltonian
including the magnetic field.  By unitarity of time
evolution, it remains a projector if the initial state describes filled
bands only. The operator $\r$ appears here, but in a commutator.  Since
$\me{\r_1}{[\mathcal{O},\r]}{\r_2} = (\r_2-\r_1)\mathcal{O}_{\r_1\r_2}$, such
expressions do not suffer from the difficulties of an ``unprotected'' $\r$, 
namely its unboundedness.  We
can only use cyclicity of the trace to the extent that this property can be
preserved.  In particular, the expression $\Tr\, \r[\rho,H]$,
which seems formally equivalent to Eq.~(\ref{eq:comm1}), poses problems,
but 
\begin{equation}
\mathbf{J}(t)= \frac{e}{\Omega} \frac{i}{\hslash}
\Tr [\rho,[\rho,\mathbf{r}]][\rho,H]
\end{equation}
does not. This expression can be derived from Eq.~(\ref{eq:comm1}) using
again the idempotency of $\rho$ ($\rho\rho=\rho$).
Using the equation of motion for the density matrix,
\begin{equation}
i\hbar\dot{\rho}(t)=[H(t),\rho(t)], \label{eq:schrodinger}
\end{equation}
and making the approximation $\rho\approx \rho^g$ on the right-hand
side at this stage,
the current becomes
\beq \label{eq:threepigs}
\mathbf{J} = \frac{e}{\Omega} \int\! d\r_1d\r_2d\r_3(\r_1-2\r_2+\r_3)
\rho^g_{\r_1\r_2}\rho^g_{\r_2\r_3}\dot{\rho}^g_{\r_3\r_1}.
\eeq

\emph{Magnetic field dependence of the current}:
The considerations given in the last subsection make the integrand
\beq
\rho^g_{\r_1\r_2}\rho^g_{\r_2\r_3}\dot{\rho}^g_{\r_3\r_1}
= \bar{\rho}^g_{\r_1\r_2}\bar{\rho}^g_{\r_2\r_3}\dot{\bar{\rho}}^g_{\r_3\r_1}
e^{-i\phi_{123}/\phi_0},
\eeq
where, again,
$\phi_{123} = \B\cdot(\r_1\times\r_2+\r_2\times\r_3+\r_3\times\r_1)/2$ is
the magnetic flux through the triangle with vertices $\r_1\r_2\r_3$ and does 
not suffer from the pathologies of $\r$ itself, which allows us to expand
$e^{-i\phi_{123}/\phi_0} = 1 - i\phi_{123}/\phi_0$ to lowest order in $B$
(recall again that the matrix elements of $\rho$ are exponentially suppressed
with distances).

Recalling our division $\bar{\rho}^g = \rho_0 + \rho'$ where $\rho'$ is of
first order in the magnetic field, Eq.~\eqref{eq:threepigs} becomes
\begin{multline} \label{eq:fullcurrent}
\mathbf{J} = \frac{e}{\Omega} \int d\r_1d\r_2d\r_3 (\r_1-2\r_2+\r_3)
\bigg[  \rho_{0\r_1\r_2} \rho_{0\r_2\r_3} \dot{\rho}'_{\r_3\r_1} \\
+ \rho'_{\r_1\r_2} \rho_{0\r_2\r_3} \dot{\rho}_{0\r_3\r_1}
+ \rho_{0\r_1\r_2} \rho'_{\r_2\r_3} \dot{\rho}_{0\r_3\r_1} \\
-i\frac{\phi_{123}}{\phi_0} 
\rho_{0\r_1\r_2} \rho_{0\r_2\r_3} \dot{\rho}_{0\r_3\r_1} \bigg]
\end{multline}
at first order.
The rest of the calculation involves substituting the expressions
for the magnetic-field dependence of $\rho^g$ obtained earlier, 
and integrating the result to obtain $\alpha$.
The energy-dependent part of $\alpha$, namely $\alpha_G$, will come from
the mixing of the occupied and unoccupied bands, Eq.~(\ref{eq:Lanswer}).
The Chern-Simons
term will come from the ``exclusion-principle--correcting'' terms,
Eq.~\eqref{eq:Xanswer}, as well as the $\phi_{123}$ term in the previous equation.

\emph{Integrating the results}:
\begin{subequations}
The four terms in the current can be collected and rearranged into
the form
\beq
\bf{J} = \bf{J}_G + \bf{J}_{CS1} + \bf{J}_{CS2}
\eeq
and integrated with respect to time as follows.
The first term in Eq.~\eqref{eq:fullcurrent} can be rewritten with 
$\rho_{0\r_1\r_2} \rho_{0\r_2\r_3} \dot{\rho}'_{\r_3\r_1}
= \d_t (\rho_{0\r_1\r_2} \rho_{0\r_2\r_3} \rho'_{\r_3\r_1})
- \dot{\rho}_{0\r_1\r_2} \rho_{0\r_2\r_3} \rho'_{\r_3\r_1}
- \rho_{0\r_1\r_2} \dot{\rho}_{0\r_2\r_3} \rho'_{\r_3\r_1}$
and combined with the next two terms to give
\begin{align}
\bf{J}_G &= \frac{e}{\Omega} \d_t \Tr [\rho_0,\r][\rho',\rho_0]  
\label{eq:totder} \\
\bf{J}_{CS1} &= - 3 \frac{e}{\Omega} \Tr \rho'[\dot{\rho}_0,[\rho_0,\r]],
\label{eq:cscurra}
\end{align}
while the final term in Eq.~\eqref{eq:fullcurrent} (\textit{i.e.}, the 
term involving $\phi_{123}$) becomes
\beq
\bf{J}_{CS2} = -i \frac{e^2}{2\hslash\Omega} B^j \epsilon_{jab} \Tr [\rho_0,\r] 
[\rho_0,r^a] [r^b,\dot{\rho}_0] + \cc \label{eq:cscurrb}
\eeq
\end{subequations}
upon rewriting 
\begin{multline}
(\r_1-2\r_2+\r_3) (\r_1\times\r_2+\r_2\times\r_3+\r_3\times\r_1) \\
= (\r_1-\r_2) [(\r_1-\r_3)\times(\r_2-\r_3)] \\
+ (\r_3-\r_2) [(\r_1-\r_2)\times(\r_1-\r_3)]. \notag
\end{multline}

The total derivative term $\bf{J}_G$ [Eq.~\eqref{eq:totder}] can be written
\beq 
J^i_G = \d_t (\alpha_G)^i_jB^j
\eeq
with $\alpha_{G}$ as given in Eq.~\eqref{eq:alphaI},
\begin{multline} \label{eq:alphaI2}
(\alpha_G)^i_j =  \frac{e^2}{\hslash}\mathrm{Re} \negthickspace
\sum_{\begin{subarray}{c}n\,\mathrm{occ}\\m\,\mathrm{unocc}\end{subarray}}
\negthickspace
\frac{ \me{n}{\d^i\P}{m} 
\me{m}{
\epsilon_{jab} \{ \d^a H, \d^b\P\} 
}{n} 
}{E_n-E_m} \\
+ 2e\,\mathrm{Im} \negthickspace
\sum_{\begin{subarray}{c}n\,\mathrm{occ}\\m\,\mathrm{unocc}\end{subarray}}
\negthickspace
\frac{ \me{n}{ \d^i\P}{m} 
\me{m}{ \d H'/\d B^j }{n} 
}{E_n-E_m},
\end{multline}
where the BZ integral and the dependence on $\k$ have been 
suppressed, and $\ket{n} = \ket{u_n}$, etc.
  This result 
follows immediately upon taking the trace in the basis of energy 
eigenstates.  Matrix elements of $[\rho_0,r^i]$ appear as 
$\not\! r^i_{\k} = \d^i\P_k$, from Eq.~\eqref{eq:notr}, and the cross-gap
matrix elements of $\rho'$ in are given in Eq.~\eqref{eq:Lanswer}.
Note that since $\bf{J}_G$ is a total time derivative, $\alpha_G$ is
uniquely defined for a given Hamiltonian (this assumes the existence
of a reference Hamiltonian with $\alpha_G = 0$, that is, the existence
of a topologically trivial, time-reversal-invariant band insulator).

In $\bf{J}_{CS2}$ [Eq.~\eqref{eq:cscurrb}],
we can replace $r \rightarrow [[r,\rho_0],\rho_0]$
in the third commutator.
This has the same cross-gap matrix elements as $r$; the interior
matrix elements do not contribute to the
trace because the other three factors,
$\dot{\rho}_0$ and 
two components of $[\rho_0,r^i]$,
have only cross-gap matrix elements.
Then
\beq \notag
\bf{J}_{CS2} = i \frac{e^2}{2\hslash\Omega} B^j \epsilon_{jab} \Tr [\rho_0,\r] 
[\rho_0,r^a] [[[\rho_0,r^b],\rho_0],\dot{\rho}_0] + \cc
\eeq 
or
\beq \notag
\bf{J}_{CS2} = -\frac{e^2}{2\hslash} B^j \epsilon_{jab} \Tr \bm{\nabla}_k \P_{\k} \d^a \P_{\k}
[ [\d^b \P_{\k},\P_{\k}],\dot{\P}_{\k}] +\cc, \\
\eeq
where an integral over $\k$ is suppressed for brevity and the trace is taken 
in the Hilbert space at $\k$.  Dropping the subscripts
$\k$ everywhere, this can be expanded and rearranged to give
\begin{multline} \tag{\ref{eq:cscurrb}'}
\bf{J}_{CS2} =
\frac{e^2}{2\hslash} B^j \epsilon_{jab} \Tr \P \{ 
[\bm{\nabla}\P,\dot{\P}]\d^a\P\d^b\P \\
+2[\dot{\P},\d^b\P][\bm{\nabla}\P,\d^a\P] \\
+3(\dot{\P}\d^b\P\d^a\P\bm{\nabla}\P 
+\d^b\P\dot{\P}\bm{\nabla}\P\d^a\P) \}.
\end{multline}
In manipulating these strings of projection operators and their derivatives,
it is very useful to realize that derivatives of projectors only have 
cross-gap matrix elements: $\P \d^a\P \P = \Q \d^a\P \Q = 0$, where 
$\Q=1-\P$ is the projector onto unoccupied bands.  This means, for example, 
that $\P (\bm{\nabla}\P)(\dot{\P}) = \P(\bm{\nabla}\P)\Q(\dot{\P})\P$.

To $\bf{J}_{CS2}$
we must add $\bf{J}_{CS1}$ [Eq.~\eqref{eq:cscurra}],
\begin{multline} \tag{\ref{eq:cscurra}'}
\bf{J}_{CS1} = 
\frac{3e^2}{2\hslash} B^j \epsilon_{jab} \Tr
(\P-\Q)[\dot{\P},\bm{\nabla}\P] \d^a\P\d^b\P \\
= \frac{3e^2}{2\hslash} B^j \epsilon_{jab} \Tr \P \{
[\dot{\P},\bm{\nabla}\P] \d^a\P\d^b\P \\
-(\dot{\P}\d^b\P\d^a\P\bm{\nabla}\P+\d^b\P\dot{\P}\bm{\nabla}\P\d^a\P) \},
\end{multline}
to get
\begin{align}
\bf{J}_{CS} &= \bf{J}_{CS1} + \bf{J}_{CS2} \notag \\
&= \frac{e^2}{\hslash} B^j \epsilon_{jab} \Tr \P \{ 
[\dot{\P},\bm{\nabla}\P]\d^a\P\d^b\P \notag \\
&\qquad+[\dot{\P},\d^b\P][\bm{\nabla}\P,\d^a\P] \}.
\end{align}
By checking the different components 
explicitly one can see that this is
\begin{multline}
\bf{J}_{CS} = 
\B \frac{e^2}{\hslash} \Tr \P \{ [\dot{\P},\d^x\P][\d^y\P,\d^z\P] \\
+ [\dot{\P},\d^y\P][\d^z\P,\d^x\P] + [\dot{\P},\d^z\P][\d^x\P,\d^y\P] \},
\end{multline}
so we get the ``topological current''
\beq
\bf{J}_{CS} = -\B \frac{e^2}{\hslash} \intbz \tr 
( \F^{tx}\F^{yz} + \F^{ty}\F^{zx} + \F^{tz}\F^{xy} ),
\eeq
where the lower-case trace ($\tr$) is only over the occupied bands, and the 
Brillouin-zone integral has been restored.

It remains only to show that $\bf{J}_{CS}$ is a total time derivative
that integrates to $\alpha_{CS}\B$.
Allowing the indices to run over
$t, x, y, z$, in that order (so that $\epsilon_{t x y z} = +1$),
\begin{align}
&\bf{J}_{CS} =-\B \frac{e^2}{8\hslash} \intbz 
\epsilon_{abcd} \tr \F^{ab}\F^{cd} \notag \\
&= -\B \frac{e^2}{2\hslash} \epsilon_{abcd} \intbz \d^a \tr  
\left( \mathcal{A}^b \partial^c \mathcal{A}^d 
- i\frac{2}{3} \mathcal{A}^b \mathcal{A}^c \mathcal{A}^d \right).
\end{align}
The derivatives with respect to $k_x,k_y,k_z$ will vanish when integrated over
the Brillouin zone assuming that $\A$ is defined smoothly and periodically over
the zone, leaving just
\beq \label{eq:JCSfinal}
\mathbf{J}_{CS} \! = \! -\B \frac{e^2}{2\hslash} \d_t \! 
\intbz \epsilon_{abc} \tr \!
\left( \mathcal{A}^a \partial^b \mathcal{A}^c 
- i\frac{2}{3} \mathcal{A}^a \mathcal{A}^b \mathcal{A}^c \right),
\eeq
where the indices now only run over $xyz$, as originally.
This obviously gives $\alpha_{CS}$ as in Eq.~\eqref{eq:alphaCS}, completing
the proof.
It must be reiterated that this integral is not always entirely trivial.  In 
particular, if the adiabatic evolution brings the crystal back to its initial 
Hamiltonian in a nontrivial way, the Brillouin zone integral need not return
to its initial value because $\A$ is not uniquely defined.  In other words, 
$\int dt \mathbf{J}_{CS}$ can be 
multivalued as a function of the Hamiltonian deformation parameters.  However,
the change can only be such that $\theta$ changes by an integer multiple of 
$2\pi$, as discussed in subsection \ref{sec:CS}.

\section{Summary}

The theoretical calculation of the magnetoelectric polarizability in insulators
presents a difficulty similar to that known well from the theory of 
polarization; both quantities suffer an inherent ambiguity in the bulk.
The magnetoelectric polarizability
adds another level of difficulty because the vector potential is unbounded and
breaks lattice translation symmetry.  However, we have developed a formalism 
that allows 
us to deal directly with a uniform magnetic field.  In the appendix, we further
show that a 
long-wavelength regularization of the vector potential together with a suitable 
generalization of the polarization (to deal with the broken crystal symmetry)
provides a (relatively) simple, though less rigorous, way to compute the 
response function.
The final expression for the
OMP rederives known results for particular model systems and topological 
insulators and completes the
picture with additional terms that have a relatively straightforward and 
intuitive interpretation.  We hope that these results and the method of their
derivation will be valuable for future work on magnetoelectric effects and
topological electronic phases.

The authors gratefully acknowledge useful discussions with  
S.~Coh, A.~Malashevich and I.~Souza.
The work was supported by the Western Institute of Nanoelectronics
(AME), DARPA OLE (AMT), NSF DMR-0804413 (JEM), and NSF DMR-0549198 (DV).

\appendix

\section{Calculating the OMP using Static Polarization}

As noted in the text, matrix elements of the operator $r$ are ill-behaved
in a basis of extended, Bloch-like states.  That problem was solved by working
with the density operator $\rho$, whose matrix elements are exponentially
suppressed with distance.  Another approach is to use a Wannier-like basis
of localized states.  In this appendix, we take this approach to present an
alternative derivation of the OMP.  

The Bloch functions $\psi_{n\k}(\r)$ of the unperturbed crystal
will evolve, under the application of a
long-wavelength magnetic field $\mathbf{A}=
\mathbf{A}_0\sin \mathbf{q}\cdot\mathbf{r}$,
into the exact energy eigenfunctions $\Psi_{n\k}(\r)$. These
no longer have a sharp crystal momentum $\k$, but may be expanded in
a perturbation series in the unperturbed $\psi_{n\k}(\r)$.
Then the analogue to the standard Wannier 
function $w_{n\bf{R}}(\r)$ for lattice vector $\bf{R}$ will be
\begin{align}
W_{n\bf{R}}(\r) &= \frac{\Omega}{\sqrt{N}} \intbz \Psi_{n\k}(\r) 
e^{-i\k\cdot\bf{R}} \notag\\
&= w_{n\bf{R}}(\r) + \delta w_{n\bf{R}}(\r),
\end{align}
where $\Omega$ is the volume of the crystal and $N$ is the number of unit 
cells.
The Wannier orbitals centered at $\mathbf{R}$ become polarized
when the magnetic field is applied, and this distortion
gives a polarization density of
\beq
\delta\mathbf{P}(\mathbf{R})=\frac{1}{\Omega}\sum_{n\,\mathrm{occ}} \me{w_{n\mathbf{R}}}{e\r}{\delta w_{n\mathbf{R}}} + \cc
\eeq
Although it is not obvious that the bulk polarization appearing
in Maxwell's equations is the same as the polarization of a set
of Wannier orbitals, this expression leads to Eq.~\eqref{eq:alphatot}.
To ensure that the Wannier orbitals are localized, we will have to 
suppose that 
each band has a vanishing Chern number,\cite{thoulesswannier} so that the phase 
of $u_{n\k}$ can be chosen so that it is a periodic function of $\k$.
In this case the unperturbed Wannier functions are localized,
and 
(though there are usually 
subtleties in defining Wannier functions in a magnetic 
fields),\cite{rashbawannier} the regularization used here
leads to localized orbitals.
Presumably these arguments can be extended to the
case where
the \emph{total} Chern number for all occupied bands
$C^{ij}=\sum_{n\, \mathrm{occ}}\int_{\mathrm{BZ}} 
d^3k \mathcal{F}^{ij}_{nn}(\k)/2\pi$ vanishes.

Here we want to take a relatively direct approach to perturbation theory in the
field, and write\cite{niugroup}
\begin{align}
\Psi_{n\k} &= \psi_{n\k} + \delta\psi_{n\k} \\
\delta\psi_{n\k} &= \frac{eB}{2iq}\sum_{l} \left[ \psi_{l\k+\q} 
\frac{ \me{u_{l\k+\q}}{v^x}{u_{n\k}} }{E_{n\k}-E_{l\k+\q}+i\epsilon} 
- (q\rightarrow-q) \right]. \notag 
\end{align}
For definiteness we take $\bf{A} = -(B/q)\sin(qy)\hat{\bf{x}}$, and the
velocity operator can be alternatively expressed as 
$v^x = \d^x H_{\k}/\hslash$, with $H_{\k}$ the Bloch Hamiltonian of the 
unperturbed crystal.

Then the first-order correction to the  dipole moment of the generalized
Wannier functions will be
\begin{multline}
\delta P^i(\mathbf{R}) = e\sum_{n\,\mathrm{occ}} 
\int d\r  \intbz \,
\left( r^i e^{i\k\cdot(\mathbf{R}-\r)}u_{n\k}^*(\r)  \right) \\ 
\times \int_{\mathrm{BZ}} \frac{d^3k'}{(2\pi)^3} \delta\psi_{n\k '}(\r)
e^{-i\k'\cdot\mathbf{R}} + \cc
\end{multline}
The position integral must be taken over the whole crystal at this point.
In the integral over $k$, $r^i$ can be converted into a $k$ derivative of
the exponential, and
then partial integration leaves a factor $-i \d_{k_i} u^*$ (the boundary
term vanishes because the Bloch function $\psi$ is strictly periodic in $k$).
Then
\begin{multline}
\delta P^i = - \frac{e^2B}{2q} 
\sum_{\begin{subarray}{c}n\,\mathrm{occ}\\ l\end{subarray}} \Bigg[
\frac{ \ip{\d^i u_{n\k} }{u_{l\k}} 
\me{u_{l\k}}{v^x}{u_{n\k-\q}} }{E_{n\k-\q}-E_{l\k}+i\epsilon}e^{i\mathbf{q}\cdot\mathbf{R}} \\
- (q\rightarrow-q) \Bigg] + \cc
\end{multline}
(From now on, we will omit the integral over $\mathbf{k}$ and
the associated factor of $(2\pi)^3$.)
Because of the variation of the magnetic field
the magnetoelectric polarization 
$\delta P^i(\bf{R})=\alpha^i_j B^j(\bf{R})$ should vary 
as $\cos\mathbf{q}\cdot\mathbf{R}$.  
The polarization seems to have both cosine and sine terms, but
the coefficient of the latter
is $-\mathbf{B}\sin(qy) C^{ix}/q$, and the vanishing of $C$ is a prerequisite
for using Wannier functions.

To lowest order in $q$ and $B$, then, the magnetoelectric response is
\begin{equation}
\alpha^i_j = -\frac{e^2}{2}\epsilon_{jab} \d_{q_b}\! 
\sum_{\begin{subarray}{c} l \\ n\,\mathrm{occ}\end{subarray}} \!
\frac{ \ip{\d^i u_{n\k} }{u_{l\k}} 
\me{u_{l\k}}{v^a}{u_{n\k-\q}} }{E_{n\k-\q}-E_{l\k}+i\epsilon} + \cc,
\end{equation}
where we have symmetrized over Landau gauges to make the expression nicer.

Switching to the shorthand $\ket{n} = \ket{u_{n\bf{k}}}$,
\begin{multline}
\alpha^i_j = \frac{e^2}{\hslash} \epsilon_{jab} \mathrm{Re} \!
\sum_{\begin{subarray}{c}n\,\mathrm{occ}\\ l\end{subarray}}
\bigg\lbrack\frac{ 
\ip{\partial^i n}{l} \me{l}{\d^a H}{\d^b n}
}{E_n - E_l+i\epsilon} \\
- \frac{ \ip{\partial^i n}{ l } \me{l}{\partial^a H}{n} \partial^b E_n
}{(E_n - E_l+i\epsilon)^2} \bigg\rbrack.
\end{multline}
Simplifying the second term of this expression makes use of the 
``Sternheimer equation''
\begin{equation}
(\partial^a H) \ket{n} = (E_n - H) \ket{\partial^a n} 
 + (\partial^a E_n) \ket{n}
\end{equation}
and the antisymmetry in the indices $a$ and $b$ to give
\begin{equation} \label{eq:compactalpha}
\alpha^i_j = \frac{e^2}{2\hslash} \epsilon_{jab}
\sum_{\begin{subarray}{c}n\,\mathrm{occ}\\ l\end{subarray}}
\frac{ \ip{\d^i n}{l} \me{l}{\d^a (H + E_n)}{\d^b n} }
{E_n - E_l +i\epsilon } + \cc
\end{equation}

Note the formal similarity to the expression for
orbital magnetization,
\begin{align}
M_j &= \frac{1}{2} \mathrm{Im} \!
\sum_{\begin{subarray}{c} n\,\mathrm{occ}\end{subarray}} \!
\epsilon_{jab} \me{\d^a n}{(H+E_n)}{\d^b n} \notag \\
&= -\frac{1}{2} \mathrm{Im} \!
\sum_{\begin{subarray}{c} n\,\mathrm{occ}\end{subarray}} \!
\epsilon_{jab} \me{n}{\d^a (H+E_n)}{\d^b n},
\end{align}
in particular the appearance of the combination 
$H+E_n$.\cite{thonhauser,niugroup}

To bring our compact expression into the form given in terms of 
$\alpha_G$ and $\alpha_{CS}$ in the main text, we need to break the 
sum over $l$ into contributions from occupied and unoccupied states.
Omitting the factor $(e^2/2\hbar)\epsilon_{jab}$ for the moment,
the sum over the occupied states takes the form

\begin{subequations}
\begin{align}
&\sum_{\begin{subarray}{c} n,l\\ \mathrm{occ}\end{subarray}} 
\frac{
\ip{\partial^i n}{l} \me{l}{\partial^a (H + E_n)}{\partial^b n}
}{E_n - E_l +i\epsilon } + \cc \notag \\
&= \sum_{\begin{subarray}{c} n,l\\ \mathrm{occ}\end{subarray}} 
\ip{\partial^i n}{l} \frac{
\me{l}{\partial^a (H - E_l)}{\partial^b n} 
+ \me{\partial^b l}{\partial^a (H - E_n)}{n}
}{E_n - E_l +i\epsilon } \notag \\
&= \sum_{\begin{subarray}{c} n,n'\\ \mathrm{occ}\end{subarray}}
\ip{\partial^i n}{n'} \ip{\partial^b n'}{\partial^a n}
\label{eq:occberry} 
\end{align}
using the antisymmetry in $a$ and $b$ and the Sternheimer
equation again.  Because the two sums are not symmetric
when we take $l$ in the unoccupied space, however, the terms do not
cancel as nicely. 
Inserting a resolution of the identity, broken into two parts, gives: 
\begin{align}
&\sum_{\begin{subarray}{c}n,n'\,\mathrm{occ}\\m\,\mathrm{unocc}\end{subarray}}
\frac{
\ip{\d^i n}{m} \me{m}{\d^a H}{n'}\ip{n'}{\d^b n}}{E_n - E_m} + \cc \notag \\
&=
\sum_{\begin{subarray}{c}n,n'\,\mathrm{occ}\\m\,\mathrm{unocc}\end{subarray}}
\ip{\d^i n}{m} \ip{m}{\d^a n'} \ip{n'}{\d^b n} + \cc
\label{eq:unoccberry} \\
&-\sum_{\begin{subarray}{c}n,n'\,\mathrm{occ}\\m\,\mathrm{unocc}\end{subarray}}
\frac{ \ip{\d^i n}{m} \ip{m}{\d^a n'} \me{n'}{\d^b H}{n} }{E_n - E_m} + \cc
\label{eq:unoccb} \\
&+\sum_{\begin{subarray}{c}n\,\mathrm{occ}\\m\,\mathrm{unocc}\end{subarray}}
\frac{ \ip{\d^i n}{m} \ip{m}{\d^a n} \d^b E_n}{E_n - E_m} + \cc
\notag
\end{align}
and
\begin{align}
\sum_{\begin{subarray}{c}n\,\mathrm{occ}\\m,m'\,\mathrm{unocc}\end{subarray}}
\frac{\ip{\d^i n}{m} \me{m}{\d^a H}{m'} \ip{m'}{\d^b n} }{E_n-E_m} + \cc 
\label{eq:unoccc} \\
+\sum_{\begin{subarray}{c}n\,\mathrm{occ}\\m\,\mathrm{unocc}\end{subarray}}
\frac{ \ip{\d^i n}{m} \ip{m}{\d^b n} \d^a E_n}{E_n - E_m} + \cc  \notag
\end{align}
\end{subequations}
The unnumbered pieces of these equations cancel by antisymmetry in $a$ 
and $b$.

Defining $\mathcal{P}$ as the projector onto occupied bands as in the text,
Eqs.~\eqref{eq:unoccb} and \eqref{eq:unoccc}
combine to give
\begin{equation}
(\alpha_G)^i_j = \frac{e^2}{2\hslash} \,
\epsilon_{jab} \negthickspace\negthickspace
\sum_{\begin{subarray}{c}n\,\mathrm{occ}\\m\,\mathrm{unocc}\end{subarray}}
\negthickspace\negthickspace
\frac{ \me{n}{\d^i\mathcal{P}}{m} \me{m}{\{\d^a H,\d^b\mathcal{P}\} }{n} 
}{E_n-E_m} + \cc,  \label{eq:nonclifford}
\end{equation}
which is equivalent to Eq.~(\ref{eq:alphaI}) upon identifying $v^a$ with 
$\d^a H$ and $\!\not\!r^i$ with $\d^i \P$.
This quantity has the crucial property that it is ``gauge invariant,'' meaning
that it can be written as a matrix trace, and hence does not change under a 
change of basis of the Hilbert space.  Of course, this property is not evident
here, where the formula makes explicit reference to energy eigenfunctions and 
their energies, but it follows from the expression in terms of a matrix given 
in Eq.~\eqref{eq:totder}.
The remainder, Eqs.~\eqref{eq:occberry} and \eqref{eq:unoccberry}, becomes
\begin{equation}
(\alpha_{\rm CS})^i_j= -\frac{e^2}{2\hslash}
\delta^i_j \epsilon_{abc} \,
\mathrm{tr} 
\left[ \mathcal{A}^a \partial^b \mathcal{A}^c 
- \frac{2i}{3} \mathcal{A}^a \mathcal{A}^b \mathcal{A}^c \right],
\end{equation}
which reproduces Eq.~\eqref{eq:alphaCS}.


\begin{thebibliography}{28}
\expandafter\ifx\csname natexlab\endcsname\relax\def\natexlab#1{#1}\fi
\expandafter\ifx\csname bibnamefont\endcsname\relax
  \def\bibnamefont#1{#1}\fi
\expandafter\ifx\csname bibfnamefont\endcsname\relax
  \def\bibfnamefont#1{#1}\fi
\expandafter\ifx\csname citenamefont\endcsname\relax
  \def\citenamefont#1{#1}\fi
\expandafter\ifx\csname url\endcsname\relax
  \def\url#1{\texttt{#1}}\fi
\expandafter\ifx\csname urlprefix\endcsname\relax\def\urlprefix{URL }\fi
\providecommand{\bibinfo}[2]{#2}
\providecommand{\eprint}[2][]{\url{#2}}

\bibitem[{\citenamefont{Spaldin and Fiebig}(2005)}]{spaldin-s05}
\bibinfo{author}{\bibfnamefont{N.~A.} \bibnamefont{Spaldin}} \bibnamefont{and}
  \bibinfo{author}{\bibfnamefont{M.}~\bibnamefont{Fiebig}},
  \bibinfo{journal}{Science} \textbf{\bibinfo{volume}{309}},
  \bibinfo{pages}{391} (\bibinfo{year}{2005}).

\bibitem[{\citenamefont{Fiebig}(2005)}]{fiebig-jpdp05}
\bibinfo{author}{\bibfnamefont{M.}~\bibnamefont{Fiebig}}, \bibinfo{journal}{J.
  Phys. D-applied Phys.} \textbf{\bibinfo{volume}{38}}, \bibinfo{pages}{R123}
  (\bibinfo{year}{2005}).

\bibitem[{\citenamefont{Qi et~al.}(2008)\citenamefont{Qi, Hughes, and
  Zhang}}]{qilong}
\bibinfo{author}{\bibfnamefont{X.-L.} \bibnamefont{Qi}},
  \bibinfo{author}{\bibfnamefont{T.~L.} \bibnamefont{Hughes}},
  \bibnamefont{and} \bibinfo{author}{\bibfnamefont{S.-C.} \bibnamefont{Zhang}},
  \bibinfo{journal}{Physical Review B} \textbf{\bibinfo{volume}{78}},
  \bibinfo{pages}{195424} (\bibinfo{year}{2008}).

\bibitem[{\citenamefont{Wilczek}(1987)}]{wilczekaxion}
\bibinfo{author}{\bibfnamefont{F.}~\bibnamefont{Wilczek}},
  \bibinfo{journal}{Phys. Rev. Lett.} \textbf{\bibinfo{volume}{58}},
  \bibinfo{pages}{1799} (\bibinfo{year}{1987}).

\bibitem[{\citenamefont{Essin et~al.}(2009)\citenamefont{Essin, Moore, and
  Vanderbilt}}]{essinomp}
\bibinfo{author}{\bibfnamefont{A.~M.} \bibnamefont{Essin}},
  \bibinfo{author}{\bibfnamefont{J.~E.} \bibnamefont{Moore}}, \bibnamefont{and}
  \bibinfo{author}{\bibfnamefont{D.}~\bibnamefont{Vanderbilt}},
  \bibinfo{journal}{Physical Review Letters} \textbf{\bibinfo{volume}{102}},
  \bibinfo{pages}{146805} (\bibinfo{year}{2009}).

\bibitem[{\citenamefont{Xiao et~al.}(2009)\citenamefont{Xiao, Shi, Clougherty,
  and Niu}}]{xiao}
\bibinfo{author}{\bibfnamefont{D.}~\bibnamefont{Xiao}},
  \bibinfo{author}{\bibfnamefont{J.}~\bibnamefont{Shi}},
  \bibinfo{author}{\bibfnamefont{D.~P.} \bibnamefont{Clougherty}},
  \bibnamefont{and} \bibinfo{author}{\bibfnamefont{Q.}~\bibnamefont{Niu}},
  \bibinfo{journal}{Phys. Rev. Lett.} \textbf{\bibinfo{volume}{102}},
  \bibinfo{pages}{087602} (\bibinfo{year}{2009}).

\bibitem[{\citenamefont{Hornreich and Shtrikman}(1968)}]{hornreichshtrikman}
\bibinfo{author}{\bibfnamefont{R.~M.}~\bibnamefont{Hornreich}} \bibnamefont{and}
  \bibinfo{author}{\bibfnamefont{S.}~\bibnamefont{Shtrikman}},
  \bibinfo{journal}{Physical Review}
  \textbf{\bibinfo{volume}{171}}, \bibinfo{pages}{1065}
  (\bibinfo{year}{1968}).

\bibitem[{\citenamefont{{Hehl} et~al.}(2008)\citenamefont{{Hehl}, {Obukhov},
  {Rivera}, and {Schmid}}}]{expttheta}
\bibinfo{author}{\bibfnamefont{F.~W.} \bibnamefont{{Hehl}}},
  \bibinfo{author}{\bibfnamefont{Y.~N.} \bibnamefont{{Obukhov}}},
  \bibinfo{author}{\bibfnamefont{J.-P.} \bibnamefont{{Rivera}}},
  \bibnamefont{and} \bibinfo{author}{\bibfnamefont{H.}~\bibnamefont{{Schmid}}},
  \bibinfo{journal}{Physics Letters A} \textbf{\bibinfo{volume}{372}},
  \bibinfo{pages}{1141} (\bibinfo{year}{2008}), \eprint{0708.2069}.

\bibitem[{foo()}]{foot1}
\bibinfo{note}{The index structure can
be used as a check, somewhat like dimensional analysis. For example,
it is a reminder that $\d P^x / \d B^y = \d M_y / \d E_x$, which has
matching indices,
rather than $\d M_x / \d E_y$.  One only has to remember
that $\mathbf{P}$ and $\mathbf{B}$ have upper indices,
while $\mathbf{M}$ and $\mathbf{E}$ have lower indices.
Vectors with upper indices correspond to directions in space.
For example $\mathbf{P}$ has
an upper index because it is given by $e\mathbf{r}$, while $\mathbf{E}$
has a lower index, in order for $\mathbf{E}\cdot\bf{d}\bm{\ell}=E_id\ell^i$ 
in Faraday's law to be balanced.  The index structure is also
useful when using ``internal coordinates'', especially in the case of 
nonorthorhombic crystals.  
One writes $\bf{P} = \sum_i P^i \bf{a}_i$, 
$\bf{B} = \sum_i B^i \bf{a}_i$, $\bf{M} = \sum_i M_i \bf{g}^i/(2\pi)$, and
$\bf{E} = \sum_i E_i \bf{g}^i/(2\pi)$, 
where $\bf{a}_i$ are the primitive lattice 
vectors and $\bf{g}^i$ are the reciprocal lattice vectors, 
$\bf{g}^i\cdot\bf{a}_j = 2\pi\delta^i_j$. This amounts
to setting the primitive vectors to 
$\hat{\mathbf{x}},\hat{\mathbf{y}},\hat{\mathbf{z}}$.}

\bibitem[{\citenamefont{Ederer and Spaldin}(2007)}]{edererspaldin}
\bibinfo{author}{\bibfnamefont{C.}~\bibnamefont{Ederer}} \bibnamefont{and}
  \bibinfo{author}{\bibfnamefont{N.~A.} \bibnamefont{Spaldin}},
  \bibinfo{journal}{Physical Review B (Condensed Matter and Materials Physics)}
  \textbf{\bibinfo{volume}{76}}, \bibinfo{pages}{214404}
  (\bibinfo{year}{2007}).

\bibitem[{\citenamefont{Batista et~al.}(2008)\citenamefont{Batista, Ortiz, and
  Aligia}}]{aligia}
\bibinfo{author}{\bibfnamefont{C.~D.} \bibnamefont{Batista}},
  \bibinfo{author}{\bibfnamefont{G.}~\bibnamefont{Ortiz}}, \bibnamefont{and}
  \bibinfo{author}{\bibfnamefont{A.~A.} \bibnamefont{Aligia}},
  \bibinfo{journal}{Physical Review Letters} \textbf{\bibinfo{volume}{101}},
  \bibinfo{pages}{077203} (\bibinfo{year}{2008}).

\bibitem[{\citenamefont{Wojde\l{} and \'I\~niguez}(2009)}]{wojdel}
\bibinfo{author}{\bibfnamefont{J.~C.} \bibnamefont{Wojde\l{}}}
  \bibnamefont{and}
  \bibinfo{author}{\bibfnamefont{J.}~\bibnamefont{\'I\~niguez}},
  \bibinfo{journal}{Phys. Rev. Lett.} \textbf{\bibinfo{volume}{103}},
  \bibinfo{pages}{267205} (\bibinfo{year}{2009}).

\bibitem[{\citenamefont{Malashevich et~al.}(to be
  published)\citenamefont{Malashevich, Coh, Souza, and
  Vanderbilt}}]{malashevichOMP}
\bibinfo{author}{\bibfnamefont{A.}~\bibnamefont{Malashevich}},
  \bibinfo{author}{\bibfnamefont{S.}~\bibnamefont{Coh}},
  \bibinfo{author}{\bibfnamefont{I.}~\bibnamefont{Souza}}, \bibnamefont{and}
  \bibinfo{author}{\bibfnamefont{D.}~\bibnamefont{Vanderbilt}}
  (\bibinfo{year}{to be published}).

\bibitem[{\citenamefont{King-Smith and Vanderbilt}(1993)}]{ksv}
\bibinfo{author}{\bibfnamefont{R.~D.} \bibnamefont{King-Smith}}
  \bibnamefont{and}
  \bibinfo{author}{\bibfnamefont{D.}~\bibnamefont{Vanderbilt}},
  \bibinfo{journal}{Phys. Rev. B} \textbf{\bibinfo{volume}{47}},
  \bibinfo{pages}{1651} (\bibinfo{year}{1993}).

\bibitem[{\citenamefont{Fu et~al.}(2007)\citenamefont{Fu, Kane, and
  Mele}}]{fu&kane&mele-2007}
\bibinfo{author}{\bibfnamefont{L.}~\bibnamefont{Fu}},
  \bibinfo{author}{\bibfnamefont{C.~L.} \bibnamefont{Kane}}, \bibnamefont{and}
  \bibinfo{author}{\bibfnamefont{E.~J.} \bibnamefont{Mele}},
  \bibinfo{journal}{Phys. Rev. Lett.} \textbf{\bibinfo{volume}{98}},
  \bibinfo{pages}{106803} (\bibinfo{year}{2007}).


\bibitem[{\citenamefont{Moore and Balents}(2007)}]{moore&balents-2006}
\bibinfo{author}{\bibfnamefont{J.~E.} \bibnamefont{Moore}} \bibnamefont{and}
  \bibinfo{author}{\bibfnamefont{L.}~\bibnamefont{Balents}},
  \bibinfo{journal}{Phys. Rev. B} \textbf{\bibinfo{volume}{75}},
  \bibinfo{pages}{121306(R)} (\bibinfo{year}{2007}).

\bibitem[{\citenamefont{Roy}()}]{rroy3d}
\bibinfo{author}{\bibfnamefont{R.}~\bibnamefont{Roy}},
  \bibinfo{journal}{Phys. Rev. B} \textbf{\bibinfo{volume}{79}},
  \bibinfo{pages}{195322} (\bibinfo{year}{2009}).


\bibitem[{foo2()}]{foot2}
\bibinfo{note}{This argument does not quite hold for the
Chern-Simons piece since the
gauge chosen for the Berry connection may not share the symmetry
of the system, 
accounting for the nontrivial value in a topological insulator.}

\bibitem[{\citenamefont{Li et~al.}(2009)\citenamefont{Li, Wang, Qi, and
  Zhang}}]{dynamicalaxion}
\bibinfo{author}{\bibfnamefont{R.}~\bibnamefont{Li}},
  \bibinfo{author}{\bibfnamefont{J.}~\bibnamefont{Wang}},
  \bibinfo{author}{\bibfnamefont{X.}~\bibnamefont{Qi}}, \bibnamefont{and}
  \bibinfo{author}{\bibfnamefont{S.-C.} \bibnamefont{Zhang}}
  (\bibinfo{year}{2009}), \eprint{arXiv:0908.1537}.


\bibitem[{\citenamefont{Avron et~al.}(1988)\citenamefont{Avron, Sadun, Segert,
  and Simon}}]{asss}
\bibinfo{author}{\bibfnamefont{J.~E.} \bibnamefont{Avron}},
  \bibinfo{author}{\bibfnamefont{L.}~\bibnamefont{Sadun}},
  \bibinfo{author}{\bibfnamefont{J.}~\bibnamefont{Segert}}, \bibnamefont{and}
  \bibinfo{author}{\bibfnamefont{B.}~\bibnamefont{Simon}},
  \bibinfo{journal}{Phys. Rev. Lett.} \textbf{\bibinfo{volume}{61}},
  \bibinfo{pages}{1329} (\bibinfo{year}{1988}).

\bibitem[{\citenamefont{Hosur et~al.}(2009)\citenamefont{Hosur, Ryu, and
  Vishwanath}}]{hosur}
\bibinfo{author}{\bibfnamefont{P.}~\bibnamefont{Hosur}},
  \bibinfo{author}{\bibfnamefont{S.}~\bibnamefont{Ryu}}, \bibnamefont{and}
  \bibinfo{author}{\bibfnamefont{A.}~\bibnamefont{Vishwanath}}
  (\bibinfo{year}{2009}), \eprint{arXiv:0908.2691}.

\bibitem[{foo3()}]{foot3}
\bibinfo{note}{Concretely, the generators of a Clifford algebra are a set of $N_c$ matrices
$\Gamma_a$ that satisfy the relation $\{\Gamma_a,\Gamma_b\} = 2\delta_{ab}$.
Then the Hamiltonians cited take the form
$H(\k)=\sum_{a=1}^{N_c}\epsilon_a(\k)\Gamma_a$. This automatically
satisfies the degeneracy and dispersion-reflection properties, 
since $H(\mathbf{k})$'s eigenvalues are $\pm\sqrt{\sum_a \epsilon_a(\k)^2}$.}

\bibitem[{\citenamefont{Guo and Franz}(2009)}]{guo}
\bibinfo{author}{\bibfnamefont{H.-M.} \bibnamefont{Guo}} \bibnamefont{and}
  \bibinfo{author}{\bibfnamefont{M.}~\bibnamefont{Franz}},
  \bibinfo{journal}{Physical Review Letters} \textbf{\bibinfo{volume}{103}},
  \bibinfo{pages}{206805} (\bibinfo{year}{2009}).

\bibitem[{\citenamefont{Levinson}(1970)}]{levinson}
\bibinfo{author}{\bibfnamefont{I.~B.} \bibnamefont{Levinson}},
  \bibinfo{journal}{Zh. Eskp. Teor. Fiz.}
  \textbf{\bibinfo{volume}{57}}, \bibinfo{pages}{660} (\bibinfo{year}{1970}),
  \bibinfo{note}{[\textit{Sov. Phys. JETP} \textbf{30}, 362 (1970)]}.

\bibitem[{foo4()}]{foot4}
\bibinfo{note}{The surface currents and
bulk currents both give a definite contribution to
the polarization when the magnetic field is varied.} 

\bibitem[{\citenamefont{Brown}(1964)}]{brown}
\bibinfo{author}{\bibfnamefont{E.}~\bibnamefont{Brown}},
  \bibinfo{journal}{Phys. Rev.} \textbf{\bibinfo{volume}{133}},
  \bibinfo{pages}{A1038} (\bibinfo{year}{1964}).

\bibitem[{\citenamefont{Zak}(1964)}]{zak}
\bibinfo{author}{\bibfnamefont{J.}~\bibnamefont{Zak}}, \bibinfo{journal}{Phys.
  Rev.} \textbf{\bibinfo{volume}{134}}, \bibinfo{pages}{A1602}
  (\bibinfo{year}{1964}).

\bibitem[{\citenamefont{Kohn}(1964)}]{kohnloc}
\bibinfo{author}{\bibfnamefont{W.}~\bibnamefont{Kohn}}, \bibinfo{journal}{Phys.
  Rev.} \textbf{\bibinfo{volume}{133}}, \bibinfo{pages}{A171}
  (\bibinfo{year}{1964}).

\bibitem[{\citenamefont{Dirac}(1931)}]{diracdensity}
\bibinfo{author}{\bibfnamefont{P.~A.~M.} \bibnamefont{Dirac}},
  \bibinfo{journal}{Proc. Cam. Phil. Soc.} \textbf{\bibinfo{volume}{27}},
  \bibinfo{pages}{240} (\bibinfo{year}{1931}), \bibinfo{note}{in \textit{The
  collected works of P.~A.~M.~Dirac, 1924-1928}, ed.~R.~H.~Dalitz, Cambridge
  University Press (1995).}

\bibitem[{\citenamefont{McWeeny}(1962)}]{mcweeny}
\bibinfo{author}{\bibfnamefont{R.}~\bibnamefont{McWeeny}},
  \bibinfo{journal}{Phys. Rev.} \textbf{\bibinfo{volume}{126}},
  \bibinfo{pages}{1028} (\bibinfo{year}{1962}).

\bibitem[{\citenamefont{Resta}(1992)}]{resta}
\bibinfo{author}{\bibfnamefont{R.}~\bibnamefont{Resta}},
  \bibinfo{journal}{Ferroelectrics} \textbf{\bibinfo{volume}{136}},
  \bibinfo{pages}{51} (\bibinfo{year}{1992}).
  
\bibitem[{\citenamefont{Thouless}(1984)}]{thoulesswannier}
\bibinfo{author}{\bibfnamefont{D.~J.} \bibnamefont{Thouless}}, 
  \bibinfo{journal}{J. Phys. C} \textbf{\bibinfo{volume}{17}}, 
  \bibinfo{pages}{L325} (\bibinfo{year}{1984}).

\bibitem[{\citenamefont{Rashba et~al.}(1997)\citenamefont{Rashba, Zhukov, and
  Efros}}]{rashbawannier}
\bibinfo{author}{\bibfnamefont{E.~I.} \bibnamefont{Rashba}},
  \bibinfo{author}{\bibfnamefont{L.~E.} \bibnamefont{Zhukov}},
  \bibnamefont{and} \bibinfo{author}{\bibfnamefont{A.~L.} \bibnamefont{Efros}},
  \bibinfo{journal}{Phys. Rev. B} \textbf{\bibinfo{volume}{55}},
  \bibinfo{pages}{5306} (\bibinfo{year}{1997}).

\bibitem[{\citenamefont{Xiao et~al.}(2005)\citenamefont{Xiao, Shi, and
  Niu}}]{niugroup}
\bibinfo{author}{\bibfnamefont{D.}~\bibnamefont{Xiao}},
  \bibinfo{author}{\bibfnamefont{J.}~\bibnamefont{Shi}}, \bibnamefont{and}
  \bibinfo{author}{\bibfnamefont{Q.}~\bibnamefont{Niu}},
  \bibinfo{journal}{Phys. Rev. Lett.} \textbf{\bibinfo{volume}{95}},
  \bibinfo{pages}{137204} (\bibinfo{year}{2005}).

\bibitem[{\citenamefont{Thonhauser et~al.}(2005)\citenamefont{Thonhauser,
  Ceresoli, Vanderbilt, and Resta}}]{thonhauser}
\bibinfo{author}{\bibfnamefont{T.}~\bibnamefont{Thonhauser}},
  \bibinfo{author}{\bibfnamefont{D.}~\bibnamefont{Ceresoli}},
  \bibinfo{author}{\bibfnamefont{D.}~\bibnamefont{Vanderbilt}},
  \bibnamefont{and} \bibinfo{author}{\bibfnamefont{R.}~\bibnamefont{Resta}},
  \bibinfo{journal}{Phys. Rev. Lett.} \textbf{\bibinfo{volume}{95}},
  \bibinfo{pages}{137205} (\bibinfo{year}{2005}).

\end{thebibliography}

\end{document}